Title:    A Model of the Roles of Essential Kinases in the Induction and Expression of Late Long-Term Potentiation

Authors:  Paul Smolen, Douglas A. Baxter, and John H. Byrne

33 pages, 7 figures, 1 table

Laboratory of Origin:

Department of Neurobiology and Anatomy
W.M. Keck Center for the Neurobiology of Learning and Memory
The University of Texas Medical School at Houston
P.O. Box 20708
Houston, TX 77225

Running Title:   Modeling Kinase Roles in L-LTP

Correspondence Address:

John H. Byrne
Department of Neurobiology and Anatomy
W.M. Keck Center for the Neurobiology of Learning and Memory
The University of Texas-Houston Medical School
P.O. Box 20708
Houston, TX 77225
Voice: (713) 500-5602
FAX: (713) 500-0623
E-mail: John.H.Byrne@uth.tmc.edu

Acknowledgements:

We thank Harel Shouval for critically reading the manuscript.

Supported by NIH grants P01 NS38310 and R01 NS50532.

## ABSTRACT

The induction of late long-term potentiation (L-LTP) involves complex interactions among second messenger cascades. To gain insights into these interactions, a mathematical model was developed for L-LTP induction in the CA1 region of the hippocampus. The differential equation-based model represents actions of protein kinase A (PKA), MAP kinase (MAPK), and CaM kinase II (CAMKII) in the vicinity of the synapse, and activation of transcription by CaM kinase IV (CAMKIV) and MAPK. L-LTP is represented by increases in a synaptic weight. Simulations suggest that steep, supralinear stimulus-response relationships between stimuli (*e.g.*, elevations in $[Ca^{2+}]$) and kinase activation are essential for translating brief stimuli into long-lasting gene activation and synaptic weight increases. Convergence of multiple kinase activities to induce L-LTP helps to generate a threshold whereby the amount of L-LTP varies steeply with the number of brief (tetanic) electrical stimuli. The model simulates tetanic, theta-burst, pairing-induced, and chemical L-LTP, as well as L-LTP due to synaptic tagging. The model also simulates inhibition of L-LTP by inhibition of MAPK, CAMKII, PKA, or CAMKIV. The model predicts results of experiments to delineate mechanisms underlying L-LTP induction and expression. For example, the cAMP antagonist RpcAMPs, which inhibits L-LTP induction, is predicted to inhibit ERK activation. The model also appears useful to clarify similarities and differences between hippocampal L-LTP and long-term synaptic strengthening in other systems.

# INTRODUCTION

Late long-term potentiation (L-LTP) in the CA1 region of the hippocampus begins approximately 1-2 hrs after electrical stimulation or after application of forskolin or BDNF. L-LTP is hypothesized to be essential for storing long-term memories (1) and has been reported to last for months (2). Because of this apparently fundamental role of L-LTP in learning, it is desirable to develop a conceptual representation of L-LTP induction and maintenance. An important component of such a representation is a mathematical model describing the role of key biochemical processes in L-LTP induction and expression. Such a model should be able to predict the outcomes of proposed experiments, and also suggest experiments to clarify aspects of L-LTP induction and expression.

Although models have been developed to describe aspects of the induction of early LTP (E-LTP) (3-5), no model of L-LTP induction and expression appears to have been developed. In contrast to E-LTP, L-LTP requires transcription and protein synthesis (6,7), and is associated with induction of numerous genes (8). L-LTP is a complex process involving the activation of numerous kinases, phosphatases, and genes. Although a complete understanding of the molecular processes underlying L-LTP is not available, we believe it is valuable to develop a model representing key processes that have been characterized experimentally. Such a model may guide further hypotheses and experimental tests, and may provide a framework for understanding core mechanisms underlying long-term synaptic change and memory.

The development of the model was based on data concerning induction of L-LTP at Schaffer collateral (SC) synapses in the hippocampal CA1 region. The SC pathway has been the focus of numerous studies because damage limited to CA1 inhibits the formation of declarative memory (9,10). Also, selective deletion of the NR1 subunit of NMDA receptors in the CA1 region impairs spatial memory and LTP (11). Experiments have suggested that a number of kinases are essential for the induction and expression of L-LTP in CA1. Therefore, the model focuses on representing the postsynaptic roles of PKA, MAPK, and other necessary kinases. The model provides insight into dynamic features, such as biochemical nonlinearities, which are essential for generating thresholds for L-LTP induction and for translating brief electrical stimuli into long-lasting synaptic changes. The model also predicts outcomes for experiments that would further delineate the mechanisms underlying L-LTP induction and expression.

# METHODS

## Model development

We developed a semi-quantitative model that incorporates proposed postsynaptic roles for protein kinase A (PKA), MAP kinase (MAPK), and CaM kinases II and IV (CAMKII, CAMKIV). Differential equations for the concentrations of kinases and substrates have an intermediate level of detail. Michaelis-Menten or first-order kinetics describe phosphorylations and dephosphorylations, and Hill functions describe phenomenologically CaM kinase activation by $Ca^{2+}$. Activation of gene expression is described phenomenologically with saturable, hyperbolic functions of the concentrations of phosphorylated transcription factors. This level of description has been used to model E-LTP induction (3,5). It keeps the number of equations manageable and promoes intuitive understanding of model dynamics.

The model does not consider stochastic fluctuations in molecule copy numbers. This simplification appears reasonable because average copy numbers are not well constrained for the species in our model. However, we note that fluctuations in molecule copy numbers would affect the rate and extent of biochemical reactions, and hence, introduce a random component into the L-LTP produced by a stimulus protocol. Fluctuations affecting the amount of L-LTP would arise not only from varying copy numbers of enzymes and substrates, but also from fluctuations in the transcription and translation of gene products essential for L-LTP. The origins and consequences of such fluctuations in gene expression have recently been reviewed (12). As more data are obtained to define the biochemical and genetic pathways responsible for L-LTP, modeling of stochasticity in these pathways will become feasible.

The model consists of 23 ordinary differential equations, and is schematized in Fig. 1. The model represents L-LTP as an increase in a synaptic weight W. Increases in W represent experimentally observed increases in excitatory postsynaptic potential (EPSP) amplitude or slope. The model does not consider L-LTP as dependent on prior E-LTP. Experimental evidence suggests these processes are independent, because application of forskolin or BDNF appears to induce a slowly developing L-LTP without E-LTP (13,14). However, essential upstream events, such as activation of specific kinases, may be common to the induction of both E-LTP and L-LTP.

Some proposed roles for CaMKII, PKA, and MAPK in L-LTP induction are as follows. CAMKII (15) and MAPK (16) phosphorylate proteins that enhance translation in the vicinity of synapses subjected to electrical stimuli. If this translation is inhibited, L-LTP is significantly impaired (17,18). Inhibition of CAMKII blocks induction of L-LTP (19). PKA phosphorylation of an unidentified substrate also appears necessary to set a "tag" at activated synapses (20). L-LTP occurs only at tagged synapses. The tag appears to allow "capture" of plasticity factors (proteins or mRNAs) produced following stimulation (21,22). In the model, activated CAMKII, PKA, and MAPK each phosphorylate a synaptic substrate, and all three phosphorylations are necessary for L-LTP.

Nuclear CAMKIV is activated by $Ca^{2+}$ influx subsequent to electrical stimuli, and can phosphorylate transcription factors such as cAMP response element binding protein (CREB) (23) and CREB binding protein (24). L-LTP induction by tetanic or theta-burst stimuli is strongly attenuated by inhibition of CAMKIV (25). In the model, elevation of nuclear $Ca^{2+}$ activates CaM kinase kinase (CAMKK). CAMKK and nuclear $Ca^{2+}$ cooperate to activate CAMKIV (Eqs. 2-3 below). CAMKIV is assumed to phosphorylate a transcription factor denoted TF-1, and this phosphorylation is necessary for L-LTP (Fig. 1).

MAPK activation leads to phosphorylation of transcription factors such as CREB and Elk-1. Elk-1 participates in induction of *zif-268* (26), a gene necessary for L-LTP (27). Induction of *Arg3.1/Arc*, necessary for L-LTP, is blocked by MAPK inhibition (28). L-LTP is blocked by MAPK inhibition (29,30). The MAPK isoforms that appear necessary for L-LTP induction are extracellular-regulated kinase (ERK) I / II (13,31). In the model, "MAPK" denotes these ERK isoforms. Active MAPK is assumed to translocate to the nucleus prior to phosphorylating a transcription factor denoted TF-2 (Fig. 1). Empirically, MAPK complexed with the CREB kinase RSK-2 translocates to the nucleus after depolarization by KCl (31). Dominant negative PKA, or inactive cAMP analogues, inhibit this translocation. The model therefore assumes PKA activity is necessary for MAPK nuclear translocation. Phosphorylation of TF-2 by MAPK and of TF-1 by CAMKII is assumed to induce expression of a representative gene essential for L-LTP. The concentration of the gene product protein is denoted [GPROD].

*Place Figure 1 near here*

Following tetani, cAMP is elevated in hippocampal slice (32, 33, see however 34). PKA is activated (33). PKA inhibition strongly attenuates tetanic L-LTP (35) and L-LTP can be induced by applying an active cAMP analogue (36). In the model, L-LTP – inducing stimuli elevate [cAMP], activating PKA. In electrically stimulated neurons, elevation of [cAMP] appears to be downstream of [$Ca^{2+}$] elevation, with [$Ca^{2+}$] elevation activating adenylyl cyclase isoforms 1 and 8 (37,38). Because data are insufficient for detailed modeling of adenylyl cyclase activation and cAMP production, the model does not describe $Ca^{2+}$ activation of cAMP production. Instead, we have simulated [cAMP] elevations with prescribed amplitudes and durations that appear consistent with the data available (discussed further below).

Each synaptic stimulus is modeled with simultaneous elevations of the concentrations of four independent variables: synaptic $Ca^{2+}$ ([$Ca^{2+}_{syn}$]), nuclear $Ca^{2+}$ concentration ([$Ca^{2+}_{nuc}$]), [cAMP], and an activation rate $k_{f,Raf}$ for Raf kinase (Eq. 5). Further details of stimulus parameters are provided in the following subsection. The concentrations, in μM, of active forms of enzymes and substrates are used as dependent variables.

In the model, 12 of the 23 dependent variables represent molecular species in the vicinity of the synapse. Stimuli activate synaptic CAMKII. Stimuli also activate synaptic Raf, which activates MAPKK, which activates MAPK. Five synaptic variables (Eqs. 5-12 below) describe the dynamics of this MAPK cascade. Activated CAMKII, MAPK, and PKA each phosphorylate a synaptic substrate, thereby generating a synaptic tag (Eqs. 15-16). These three synaptic tag substrates are dependent variables (Eq. 16). [GPROD], the concentration of a gene product necessary for L-LTP, is also a synaptic variable. The remaining two synaptic dependent variables are the synaptic weight W and the concentration of a protein P, which limits increase of W (Eqs. 18-19). Stimuli also activate PKA *via* cAMP. Concentrations of active PKA and of cAMP are each represented by an averaged (lumped) variable that does not distinguish between the synapse and the soma. To allow for coupling of stimuli to activation of nuclear MAPK, the model also represents activation of a somatic Raf – MAPK cascade. Five somatic dependent variables describe this cascade. The model assumes that identical equations and parameters describe the somatic and synaptic MAPK cascades, except for two terms describing nuclear import and export of somatic MAPK (Eq. 13). The remaining five dependent variables are nuclear. These are the concentrations of active nuclear MAPK, CAMKK, and CAMKIV, and the degrees of phosphorylation of the transcription factors TF-1 and TF-2.

For simplicity, a minimal representation of the coupling between synaptic, somatic, and nuclear processes is adopted. Phosphorylation of TF-1 and TF-2 is assumed to directly increase the rate of synthesis of the synaptic gene product GPROD (Eq. 17). Therefore, the transport of GPROD from nucleus to synapse is not modeled. Activated somatic MAPK is transported into the nucleus (Eqs. 13-14), and the active nuclear MAPK can then phosphorylate TF-2. No other coupling between cellular compartments is represented.

Activation of CAMKII by synaptic $Ca^{2+}$ is described by the following differential equation, which uses a Hill function of $[Ca^{2+}_{syn}]$,

$$\frac{d[CAMKII_{act}]}{dt} = k_{act1} \frac{\left[Ca^{2+}_{syn}\right]^4}{\left[Ca^{2+}_{syn}\right]^4 + K_{syn}^{\phantom{syn}4}} - k_{deact1}[CAMKII_{act}] \qquad 1)$$

Equations similar to Eq. 1 describe the activation of CaM kinase kinase (CAMKK) and CAMKIV by elevated nuclear $Ca^{2+}$. The equation for $[CAMKK_{act}]$ is,

$$\frac{d[CAMKK_{act}]}{dt} = k_{act2} \frac{\left[Ca^{2+}_{nuc}\right]^4}{\left[Ca^{2+}_{nuc}\right]^4 + K_{nuc}^{\phantom{nuc}4}} - k_{deact2}[CAMKK_{act}] \qquad 2)$$

For $[CAMKIV_{act}]$, the rate of activation is also proportional to CAMKK activity, yielding

$$\frac{d[CAMKIV_{act}]}{dt} = k_{act3}[CAMKK_{act}] \frac{\left[Ca^{2+}_{nuc}\right]^4}{\left[Ca^{2+}_{nuc}\right]^4 + K_{nuc}^{\phantom{nuc}4}}$$
$$- k_{deact3}[CAMKIV_{act}] \qquad 3)$$

In Eqs. 1-3, the Hill coefficients are given standard values of 4. These Hill functions constitute a minimal representation of the activation of CaM kinases by calmodulin (CaM), because four $Ca^{2+}$ ions bind cooperatively to CaM and CaM-$Ca_4$ activates CaM kinases. For CAMKII, data suggest a steep $[Ca^{2+}]$ dependence that can be characterized by a Hill coefficient $\geq 4$ (39). Use of a Hill coefficient greater than 4 for CAMKII did not significantly affect the simulations discussed below. For CAMKIV activity, a steep $[Ca^{2+}]$ dependence is likely given CaM-$Ca_4$'s obligatory binding to both CAMKIV and CAMKK.

Electrical or chemical stimuli are also assumed to elevate $[cAMP]$. For cAMP to activate PKA, two cAMP molecules must bind cooperatively to the regulatory (R) subunit of the PKA holoenzyme (40). Therefore, one qualitative representation of PKA activation assumes the activation rate is a Hill function of the second power of $[cAMP]$. The level of active PKA, $[PKA_{act}]$, is also assumed to undergo first-order decay due to deactivation. The resulting differential equation is,

$$\frac{d[PKA_{act}]}{dt} = \left\{ f([cAMP]) - [PKA_{act}] \right\} \Big/ \tau_{PKA}$$

with

$$f([cAMP]) = \frac{[cAMP]^2}{K_{camp}^{\phantom{camp}2} + [cAMP]^2} \qquad 4)$$

As noted above, $[cAMP]$ and $[PKA_{act}]$ are averaged variables that represent both synaptic and somatic cAMP levels and PKA activities. For fixed $[cAMP]$, Eq. 4 yields a steady-state $[PKA_{act}]$ equal to the Hill function of $[cAMP]^2$.

Stimuli that induce L-LTP are assumed to phosphorylate and activate the first kinase in a synaptic MAPK cascade, commonly Raf-1 or B-Raf in neurons (41,42). Active Raf phosphorylates MAP kinase kinase (MAPKK) twice, activating MAPKK. MAPKK then phosphorylates MAPK twice, activating MAPK. These phosphorylations can be described by the following differential equations (43)

$$\frac{d\left[Raf\right]}{dt} \; = \; -k_{f,Raf}\cdot\left[Raf\right] + k_{b,Raf}\cdot\left[Raf^p\right] \hspace{2cm} 5)$$

$$\left[Raf^p\right] = \left[Raf\right]_{tot} - \left[Raf\right] \hspace{2cm} 6)$$

$$\frac{d\left[MAPKK\right]}{dt} \; = \; -k_{f,MAPKK}\cdot\left[Raf^p\right]\cdot\frac{\left[MAPKK\right]}{\left[MAPKK\right] + K_{MKK}}$$
$$+ \; k_{b,\,MAPKK}\cdot\frac{\left[MAPKK^p\right]}{\left[MAPKK^p\right] + K_{MKK}} \hspace{1cm} 7)$$

$$\frac{d\left[MAPKK^{pp}\right]}{dt} \; = \; k_{f,MAPKK}\cdot\left[Raf^p\right]\cdot\frac{\left[MAPKK^p\right]}{\left[MAPKK^p\right] + K_{MKK}}$$
$$- \; k_{b,\,MAPKK}\cdot\frac{\left[MAPKK^{pp}\right]}{\left[MAPKK^{pp}\right] + K_{MKK}} \hspace{1cm} 8)$$

$$\left[MAPKK^p\right] = \left[MAPKK\right]_{tot} - \left[MAPKK\right] - \left[MAPKK^{pp}\right] \hspace{1cm} 9)$$

$$\frac{d\left[MAPK\right]}{dt} \; = \; -k_{f,MAPK}\cdot\left[MAPKK^{pp}\right]\cdot\frac{\left[MAPK\right]}{\left[MAPK\right] + K_{MK}}$$
$$+ \; k_{b,\,MAPK}\cdot\frac{\left[MAPK^p\right]}{\left[MAPK^p\right] + K_{MK}} \hspace{1cm} 10)$$

$$\frac{d\left[MAPK^{pp}\right]}{dt} = k_{f,MAPK}\cdot\left[MAPKK^{pp}\right]\cdot\frac{\left[MAPK^p\right]}{\left[MAPK^p\right] + K_{MK}}$$
$$- \; k_{b,\,MAPK}\cdot\frac{\left[MAPK^{pp}\right]}{\left[MAPK^{pp}\right] + K_{MK}} \hspace{1cm} 11)$$

$$[MAPK^p] = [MAPK]_{tot} - [MAPK] - [MAPK^{pp}] \qquad 12)$$

The concentration of active MAPK, $[MAPK_{act}]$, is assumed equal to $[MAPK^{pp}]$. In Eq. 5, $k_{f,Raf}$ is assigned a small positive value in the absence of stimulation, yielding some basal MAPK activation. Basal ERK activity has been observed in hippocampal neurons (44). L-LTP – inducing stimuli briefly elevate $k_{f,Raf}$.

Activated MAPK can undergo PKA-driven nuclear translocation (31). To model nuclear MAPK activity, it is necessary to represent stimulus-induced activation of a somatic MAPK cascade and nuclear translocation of somatic MAPK. To represent somatic Raf and MAPKK activation, equations identical to Eqs. 5-9 are used. Kinetic parameters (Table I) and stimulus-induced Raf activation are assumed identical for the somatic and synaptic cascades. Current data do not allow differences between somatic and synaptic parameters to be well specified, thus our assumption of identity appears reasonable for a qualitative representation. Parameter alterations during simulations (*e.g.* inhibition of MAPKK activation) are applied identically to the synaptic and somatic MAPK cascades. To represent somatic MAPK dynamics, equations identical to Eqs. 10-12 were used, except that the differential equation for somatic $[MAPK^{pp}]$ incorporates nuclear import and export. The synaptic and somatic MAPK cascades are assumed not to interact due to their spatial separation.

The model assumes that activated somatic MAPK, $MAPK_{soma}^{pp}$, undergoes nuclear import at a rate proportional to PKA activity ($[PKA_{act}]$). The concentration of <u>active</u> nuclear MAPK is denoted $[MAPK_{nuc}]$. Nuclear export of MAPK is modeled as a first-order process. The above assumptions are expressed by the following differential equations for $[MAPK_{nuc}]$ and $[MAPK_{soma}^{pp}]$,

$$\frac{d[MAPK_{nuc}]}{dt} = k_{nuc}[PKA_{act}]\left[MAPK_{soma}^{pp}\right] - k_{cyt}\left[MAPK_{nuc}\right] \qquad 13)$$

$$\frac{d\left[MAPK_{soma}^{pp}\right]}{dt} = k_{f,MAPK} \times \left[MAPKK_{soma}^{pp}\right] \times \frac{\left[MAPK_{soma}^{p}\right]}{\left[MAPK_{soma}^{p}\right]+K_{MK}}$$

$$- k_{b,MAPK} \times \frac{\left[MAPK_{soma}^{pp}\right]}{\left[MAPK_{soma}^{pp}\right]+K_{MK}} - k_{nuc}[PKA_{act}]\left[MAPK_{soma}^{pp}\right]+k_{cyt}\left[MAPK_{nuc}\right] \qquad 14)$$

The synaptic tag that "marks" synapses for L-LTP involves covalent modifications that place a synapse in a labile state capable of "capturing" plasticity factors (proteins or mRNAs) and incorporating them to increase synaptic strength. PKA appears responsible for at least one of these modifications (20). However, other kinases are needed to induce the labile state. Postsynaptic CAMKII activity is required for L-LTP. Synaptic MAPK is also likely to contribute by phosphorylating proteins that enhance local translation (45,16). Therefore, setting a synaptic tag appears to require CAMKII, MAPK, and PKA. In the model (Fig. 1), tagging is assumed to require phosphorylation of three substrates; Tag-1, Tag-2, and Tag-3. These species are respectively substrates of CAMKII, PKA, and synaptic MAPK. A molecular candidate for Tag-1 is the mRNA translation factor termed cytoplasmic polyadenylation element binding protein (CPEB), because CAMKII stimulates protein synthesis through phosphorylation of CPEB (46). The fractions of the kinase substrates that are phosphorylated are represented as deterministic variables denoted Tag-1P – Tag-3P. Their values range from 0 to 1. For simplicity, the model assumes that these three phosphorylations by different kinases are independent. With this assumption, the amount of synaptic tag, denoted as TAG, can be represented as the product of the phosphorylated fractions,

$$\text{TAG} = \text{Tag-1P} \times \text{Tag-2P} \times \text{Tag-3P} \qquad 15)$$

Phosphorylations of the transcription factors TF-1 and TF-2 are also described as fractions varying from 0 to 1. Because the model assumes the Tag phosphorylations and the TF phosphorylations are all independent from each other, the differential equations governing phosphorylation of Tag-1 – Tag-3, TF-1, and TF-2 each contain only one of these variables. These equations are all analogous to the equation for the phosphorylation of Tag-1,

$$\frac{d\left(\text{Tag-1P}\right)}{dt} = \left[\text{CAMKII}_{act}\right] k_{phos1} \left[1.0 - \left(\text{Tag-1P}\right)\right] - k_{deph1}\left(\text{Tag-1P}\right) \qquad 16)$$

Phosphorylation and dephosphorylation rate constants for Tag-2, Tag-3, TF-1, and TF-2 are respectively denoted $k_{phos2} - k_{phos5}$ and $k_{deph2} - k_{deph5}$. The kinase activities governing these phosphorylations are respectively $[\text{PKA}_{act}]$, $[\text{MAPK}_{act}]$, $[\text{CAMKIV}_{act}]$, and $[\text{MAPK}_{nuc}]$.

The rate of synthesis of the gene product GPROD that is incorporated into tagged synapses is a saturable function of the degrees of phosphorylation of TF-1 and TF-2. [GPROD] also undergoes first-order decay, yielding the following differential equation for [GPROD],

$$\frac{d\left[\text{GPROD}\right]}{dt} = k_{syn} \frac{\text{TF-1-Phos}}{\left(\text{TF-1-Phos}\right)+K_1} \frac{\text{TF-2-Phos}}{\left(\text{TF-2-Phos}\right)+K_2}$$
$$+ k_{synbas} - k_{deg}\left[\text{GPROD}\right] \qquad 17)$$

Equation 17 includes a constitutive, unstimulated GPROD synthesis rate $k_{synbas}$.

A synaptic weight W represents changes in synaptic strength due to L-LTP induction, which requires both synaptic tagging and increased gene product level. The rate of increase of W is assumed proportional to the overlap, or product, of the tag with the gene product level. As discussed further below, the increase in W is assumed to be limited by the availability, for synaptic incorporation, of another precursor molecular species denoted P. These considerations yield the following differential equation,

$$\frac{dW}{dt} = k_W\left(\text{TAG}\right)\left[\text{GPROD}\right]\frac{[P]}{[P]+K_P} - W\big/\tau_W \qquad 18)$$

Equation 18 with [P] fixed implies that W would increase indefinitely as stimulus number or duration was increased. In simulations of tetanic L-LTP with [P] fixed, the amount of L-LTP increased steeply with tetani so that 8 tetani produced a 50-fold greater W increase than 3 tetani. To remove this implausible L-LTP increase, a saturation mechanism was included, so that more than 4 tetani no longer enhanced L-LTP substantially. Because current data do not appear to demonstrate saturation of kinase activation, we used a hypothetical mechanism, in which the level of available precursor P in Eq. 18 is assumed to decrease when W increased, corresponding to incorporation of P into strengthened synapses. This assumption is expressed in the following differential equation,

$$\frac{d[P]}{dt} = V_P - k_W(TAG)[GPROD]\frac{[P]}{[P]+K_P} - [P]/\tau_P \qquad 19)$$

In Eq. 19, the rate of synthesis of P equals the parameter $V_P$. Stimuli that elevate TAG and [GPROD] decrease [P] *via* the second term on the right-hand side, which represents incorporation of P into a strengthened synapse. Eqs. 18 and 19 ensure multiple tetani increases W only to the extent allowed by depletion of available P.

Phosphatidylinositol-3-kinase (PI3K) inhibition has been reported to block the expression of E-LTP (47), but these experiments were of insufficient duration to establish the role of PI3K in L-LTP. Therefore, the model does not currently represent dynamics of PI3K activity (but see Discussion). PI3K can activate the atypical protein kinase C isoform termed PKM / PKCζ (48,49). However, this pathway has not been well studied in neurons.

Data do not generally exist to accurately determine concentrations of active enzymes in neurons. Therefore, we were not able to quantitatively fit time courses of concentrations or enzyme activities to data. However, we did obtain semi-quantitative constraints from estimates of Bhalla and Iyengar for concentrations of MAPK, PKA, PKC, CAMKII, MAPKK, and Raf (Reference 5, see www.mssm.edu/labs/iyengar/ssupplementary_materials.shtml, henceforth denoted B&I). We set [MAPK]$_{tot}$, [MAPKK]$_{tot}$, and [Raf]$_{tot}$ to 0.25 μM, close to the B&I estimates of 0.36 μM, 0.18 μM, and 0.2 μM respectively. Active PKA, [PKA$_{act}$], peaks at 0.6 μM during simulated forskolin application, whereas B&I estimate 0.5 μM for the $R_2C_2$ tetramer. This tetramer is ~80% of total PKA in unstimulated cells (50). The simulated peak concentration of active CAMKII is 7.9 μM (simulation of Fig. 3A before scaling output). The B&I estimate of total CAMKII is 70 μM. Thus, the simulated peak concentration of active CAMKII is 11% of the estimated total. Simulated peak concentrations of active CAMKIV and CAMKK due to tetanus are 0.05 and 0.1 μM, respectively. These values are ~5-10% of the total CAMKIV and CAMKK concentrations, which B&I estimate at $0.5 - 1$ μM. The qualitative simulation results discussed below (Figs. 3-7) are not sensitive to these parameter values. The concentration time course of any variable can be rescaled with preservation of the model dynamics, if kinetic rate constants relating that variable to others are rescaled. For example, [CAMKII$_{act}$] can be doubled by doubling $k_{act}$ in Eq. 1, but if $k_{phos}$ in Eq. 16 is also halved, the rate of the phosphorylation catalyzed by CAMKII stays the same and the dynamics are unchanged.

Standard parameter values are given in Table I. These values were used in all simulations except as noted below. Table I does not include values for the independent variables describing stimulus input, which are given below.

*Place Table I near here*

## Simulation of L-LTP – inducing stimuli

Stimulation protocols (Fig. 2) lead to elevation of $[Ca^{2+}]$ and $[cAMP]$ and activation of the MAPK signaling cascade. Details of $Ca^{2+}$ dynamics were not modeled, given that the model of Fig. 1 is a qualitative representation of the roles of kinases essential for L-LTP induction. Instead, the $Ca^{2+}$ response to stimuli was modeled in the simplest plausible manner. Two independent variables were used, synaptic $[Ca^{2+}]$ and nuclear $[Ca^{2+}]$. Basal $[Ca^{2+}_{syn}]$ and $[Ca^{2+}_{nuc}]$ values were 40 nM. Tetanic and theta-burst stimuli were modeled as square-wave increases in $[Ca^{2+}_{syn}]$ and $[Ca^{2+}_{nuc}]$. For tetanic stimuli, three tetani were usually simulated, with an inter-stimulus interval of 5 min (Fig. 2A). Each 1-sec, 100 Hz tetanus was simulated as a 3-sec increase of synaptic $Ca^{2+}$ to 1 μM and nuclear $Ca^{2+}$ to 500 nM. A similar duration of $Ca^{2+}$ elevation is suggested by data. One study (51) used a photolabile $Ca^{2+}$ buffer to terminate postsynaptic $Ca^{2+}$ elevation after tetani. Delaying buffer photolysis for 2.5 sec did not attenuate LTP, whereas photolysis within 2 sec inhibited LTP. More recent imaging data also suggest a time constant of 1–3 sec for decay of $Ca^{2+}$ transients after tetanus (52) although another study (53) found a more rapid decay and a higher peak $[Ca^{2+}]$ (4-6 μM). Changes in $[cAMP]$ and MAPK activity produced by the simulated tetani and other protocols are discussed below.

In theta-burst stimulus protocols, 10-12 bursts of four 100 Hz pulses are typically delivered 200 msec apart (total duration ∼ 2.2 sec) (*e.g.* reference 14). This protocol was simulated with a 5-sec square-wave increase in $[Ca^{2+}_{syn}]$ (to 1 μM) and $[Ca^{2+}_{nuc}]$ (to 500 nM) (Fig. 2B). We also simulated L-LTP induction with the pairing protocol used in (54), which uses multiple pairings of a single action potential (AP) in the potentiated synaptic pathway with a burst stimulus in a second pathway. Sixty bursts of three 100 Hz AP's are given 5 sec apart, for a total duration of 5 min. We modeled each AP-burst pairing with a brief (1.2 sec) elevation of $[Ca^{2+}_{syn}]$ (to 400 nM) and $[Ca^{2+}_{nuc}]$ (to 180 nM) (Fig. 2C).

As noted above, the kinetics of cAMP production and its activation by $Ca^{2+}$ have not been well characterized. Therefore, we assumed each tetanus or theta-burst induced a prescribed, square-wave elevation of $[cAMP]$. Observations suggest that the time for $[cAMP]$ to return to basal levels after stimulation is ∼ 1-2 min (55,56). Therefore, we assumed $[cAMP]$ remained elevated for 1 min during and after stimulation. The pairing protocol (54) was assumed to elevate $[cAMP]$ for 6 min (protocol duration + 1 min). Specific values for $[cAMP]$ were 0.05 μM (basal), 0.15 μM (tetanic), 0.35 μM (theta-burst), and 0.15 μM (pairing).

Neuronal MAPK can be activated by $Ca^{2+}$ elevation acting *via* CaM kinase I (57) or by cAMP elevation (58-60) or by a $Ca^{2+}$-independent pathway involving mGluR5 (61). Raf activation is the convergence point for these mechanisms of MAPK cascade activation. Rather than modeling these complexities in detail, we assumed each tetanus or theta-burst increased the rate constant $k_{f,Raf}$ for synaptic and somatic Raf phosphorylation and activation (Eq. 5). In the absence of detailed data, we assumed a square-wave increase lasting for 1 min for tetanic and theta-burst stimuli and 6 min for the pairing protocol. Values for $k_{f,Raf}$ were 0.0075 $min^{-1}$ (basal), 0.16 $min^{-1}$ (tetanus), 0.41 $min^{-1}$ (theta-burst), and 0.16 $min^{-1}$ (pairing). As discussed above, identical equations and $k_{f,Raf}$ values describe synaptic and somatic Raf activation. $k_{f,Raf}$ and $[cAMP]$ elevations needed to be higher for theta-bursts than for tetani, so that similar peak MAPK activation, gene induction, and L-LTP resulted after one theta-burst *vs.* after three tetani.

We also simulated "chem-LTP", in which application of forskolin or BDNF activates PKA and MAPK (62,14). Typical experimental applications last ~ 30 min. For 30 min, $k_{f,Raf}$ was elevated to 0.3 $min^{-1}$ and [cAMP] was elevated to 0.4 µM. Synaptic and nuclear [$Ca^{2+}$] were slightly elevated, by 60 nM, for 30 min. Data suggests neuronal [$Ca^{2+}$] is elevated by exposure to forskolin or BDNF (63,64).

## Modeling Synaptic Tagging and Heterosynaptic L-LTP

The model was extended to simulate sequential tetanic stimulation of two synapses, A and B, with GPROD synthesis blocked during tetanization of synapse B (Fig. 7 below). Experimentally, if protein synthesis is blocked during tetanization of synapse B, L-LTP of synapse B still results (22). The synaptic tag hypothesis (21,22) suggests that the tetanus to synapse B activates synaptic kinases and phosphorylates tag substrates. L-LTP results because gene expression and protein synthesis was induced by the prior tetani at synapse A. The necessary proteins are then "captured" by the tagged synapse B.

The model extension was carried out as follows. The differential equations for the 12 dependent synaptic variables were duplicated (Eqs. 1, 4, 5-12, 16, 18-19) and the synaptic tag was duplicated (Eq. 15). The independent stimulus variables [cAMP], $k_{f,Raf}$, and [$Ca^{2+}_{syn}$] were duplicated for synapse B. Tetanus of either synapse was simulated by brief elevations of these stimulus variables at only the tetanized synapse. Tetanus of either synapse elevates [$Ca^{2+}_{nuc}$], activating CAMKIV, and also elevates somatic $k_{f,Raf}$, activating the somatic MAPK cascade. PKA is also activated, enhancing MAPK nuclear translocation. For all stimulus variables, the basal and elevated levels are identical for stimulus of synapses A and B. These values are as given above (preceding subsection).

The only coupling between synapses A and B is *via* the nucleus. Stimulation of either synapse induces activation of the nuclear kinases (CAMKK, CAMKIV, and MAPK) and elevation of the level of GPROD at both synapses. In Fig. 7, to simulate the experimental block of protein synthesis by anisomycin, GPROD synthesis is blocked ($k_{syn}$ and $k_{synbas}$ in Eq. 17 are set to zero) during and after tetanus of synapse B.

This extension of the model simulates tagging and L-LTP of synapse B when GPROD synthesis is blocked during and after tetanus of synapse B (Fig. 7). However, to simulate more general stimulus protocols with multiple synapses, it would be necessary to represent cumulative activation of somatic PKA, which drives nuclear import of active MAPK. Separate variables would be required to represent PKA activity at the soma and at each synapse.

## Numerical methods

The forward Euler method was used for integration, with a time step of 15 msec. Simulations verified that further reductions in the time step did not significantly improve the accuracy of the results illustrated in Figs. 3-7. To further verify accuracy, the simulations of Figs. 3 and 6 were repeated using the second-order Runge-Kutta integration method (65). No significant differences in the time courses of the model variables resulted.

Initial values for the model variables were as follows. Somatic and synaptic [Raf], [MAPKK], and [MAPK] were respectively set to 0.5*[Raf]$_{tot}$, 0.5*[MAPKK]$_{tot}$, and 0.5*[MAPK]$_{tot}$. [MAPK$_{nuc}$] was set to 0.2*[MAPK]$_{tot}$. The remaining 16 dependent variables were set to 0.001. [Ca$^{2+}_{syn}$] and [Ca$^{2+}_{nuc}$] were set to 40 nM. To allow the model to reach equilibrium, simulations were run for at least four simulated days prior to L-LTP induction. During the equilibration simulation only, in order to ensure complete equilibration, the variables with the slowest time constants (W and [P]) were set equal to their steady-state values as determined by the other model variables. We verified that integration for even longer times did not alter the equilibrium state. The model was programmed in Java and simulated on Pentium 3 microcomputers. Programs are available upon request.

To allow concurrent visualization of variables of different magnitudes, amplitude scaling factors were applied when plotting simulation results (Figs. 3-7), as follows. The time courses of [Raf$^p$] and [CAMKIV$_{act}$] were vertically scaled (multiplied) by 10. [CAMKII$_{act}$] was vertically scaled by 0.1. MAPK species concentrations were scaled by 5.0. TAG was scaled by 110. [GPROD] was scaled by 0.4. In Figs. 3-7, the variables representing enzyme concentrations and the variables [P] and [GPROD] have units of μM. The other variables, such as W and TAG, are non-dimensional.

# RESULTS

## Simulation of tetanic L-LTP

Tetanic L-LTP induction was simulated by applying three tetani, with an interstimulus interval (ISI) of 5 min (Fig. 2A). After the tetani, CAMKII remains active for ~ 5 min and CAMKIV for ~ 45 min (Fig. 3A). The time required for decay of CAMKIV activity is similar to data (23). PKA activity increases by ~100% during L-LTP induction, which is consistent with data (33). Simulated synaptic and somatic MAPK activity ([MAPK$_{act}$] and [ MAPK$_{som}^{pp}$ ]) both last ~2 hr (Fig. 3B). Data concerning the duration of MAPK activity are contradictory. One recent study suggests MAPK remains phosphorylated, and presumably active, for at least 8 hr after tetanus (66). However, earlier studies (67,30) suggest a much briefer activation of ~30 min. Because long-lasting MAPK activity could regulate transcription and other processes involved in L-LTP, we suggest further experimental study of MAPK kinetics is warranted. Simulated basal [MAPK$_{act}$] is approximately 15% of peak [MAPK$_{act}$]. L-LTP induction nears completion in ~2 hr (Fig. 3D, time course of W). Similarly, induction of L-LTP with BDNF (bypassing E-LTP) requires ~2 hr (13). In Fig. 3D, W increases by 145%. This amplitude is similar to the EPSP increase observed after three or four 1 sec, 100 Hz tetani (68,30).

*Place Figure 3 near here*

In Fig. 3C, the synaptic tag variable and gene product level are both plotted to illustrate their overlap. Equation 18, describing the increase in W, represents the amount of L-LTP as proportional to this overlap. The time course of [P] is illustrated in Fig. 3D. In the model, P is assumed to limit the amount of L-LTP generated by prolonged stimuli, with synaptic incorporation of P both increasing W and diminishing [P] (Eqs. 18-19). With the parameters of Fig. 3, simulation of four tetani does generate a significantly greater elevation of W (174%). However, simulation of 10 tetani causes only a slightly greater W elevation (186%), because [P] declines to ~ 0.

**Effects of supralinear stimulus-response relationships**

The model incorporates three supralinear stimulus-response relationships. First, the rates of activation of CAMKII, CAMKK, and CAMKIV are determined by nonlinear Hill functions of $[Ca^{2+}]$. Second, active Raf phosphorylates MAP kinase kinase (MAPKK) twice. MAPKK-PP then phosphorylates MAPK twice. Only MAPK-PP phosphorylates MAPK substrates at a significant rate. These multiple phosphorylations of MAPKK and MAPK generate supralinearity in the output of the MAPK cascade (MAPK activity) as a function of the input (the rate of Raf activation) (69). Third, multiple kinase activities converge to increase W. The rate of increase of W is proportional to gene product concentration ([GPROD]) and to the synaptic tag (TAG). The rate of GPROD formation is proportional to phosphorylation of two transcription factors and therefore to the activities of CAMKIV and nuclear MAPK (with saturation at high activities). TAG is proportional to the phosphorylation of three sites and therefore to the activities of synaptic CAMKII, MAPK, and PKA. Thus, if the activities of CAMKII, CAMKIV, PKA, and MAPK are doubled, the rate of increase of W can increase by up to 16-fold.

Empirically, a ~2-3 sec, 10-20 fold elevation of $Ca^{2+}$ (from ~40 nM basal levels to ~1 μM in the vicinity of tetanized synapses, or ~300 nM at the nucleus) suffices for long-lasting gene induction (induction of *Arg3.1/Arc* and other LTP-associated genes lasts > 30 min) (28,8). Such amplification of a brief input into a long-lasting output requires steep, supralinear relationships of input ($Ca^{2+}$ elevation) to output (gene induction or synaptic weight changes). Without supralinearity, a 20-fold elevation of $[Ca^{2+}]$ lasting for 3 sec would drive only a negligible increase in a variable such as gene product concentration. The much longer time constant of the latter variable would almost completely damp its response to the brief stimulus.

To quantify the effect of the three supralinearities discussed above, we repeated the simulation of Fig. 3 in three different ways, with supralinearity reduced as follows. Case I: the $[Ca^{2+}]$ Hill coefficient in Eq. 1 was reduced to 1. Case 2: only single phosphorylations of MAPK and MAPKK were assumed to occur. Case 3: convergence of multiple kinases was reduced by elimination of the CAMKIV substrate TF-1. The basal synthesis rate of P was elevated tenfold in cases 1-3 to ensure decrease of L-LTP was not due to depletion of P. L-LTP (the increase in W) was reduced to 5.8% (Case 1), 97% (Case 2), and 5.5% (Case 3), compared to 145% in Fig. 3D. Therefore, high $[Ca^{2+}]$ Hill coefficients and convergence of multiple kinases (Cases 1 and 3) contribute substantially to simulated L-LTP. The double phosphorylations of MAPKK and MAPK (Case 2) contribute considerably less.

Supralinear stimulus-response relationships also cause simulated L-LTP to exhibit threshold behavior. In Fig. 3D, W increases by 145% following three tetani. If only two tetani are simulated, the amount of L-LTP decreases by more than half, and if only one tetanus is simulated L-LTP decreases by a further 80%. Such threshold dynamics may help explain the experimental requirement of 3-4 tetani for the reliable induction of L-LTP.

**Sensitivity of L-LTP induction to parameters and stimulus pattern**

Biochemical and genetic systems are commonly observed to be robust to significant changes in the values of parameters, such as mutations that alter enzyme activities. Therefore, a plausible model of L-LTP induction should be robust, such that simulated stimulus responses should not exhibit very high sensitivity to small changes in parameter values. However, it is also desirable to use modeling to predict parameters to which L-LTP induction may be most sensitive. Some of these "high-sensitivity" parameters could function as physiological control parameters to regulate LTP induction, and might serve as targets for pharmacological intervention to augment L-LTP and memory.

A standard method defines a set of relative sensitivities $S_i$, with the index i ranging over all parameters $p_i$ [70,71]. Let R denote the amplitude of a simulated stimulus response. For each $p_i$, a small change is made, and the resulting change in R is determined. The relative sensitivity $S_i$ is then defined as the <u>relative,</u> or <u>fractional,</u> change in R divided by the <u>relative</u> change in $p_i$,

$$S_i = \frac{\Delta R / R}{\Delta p_i / p_i} \qquad (20)$$

We chose R to be the magnitude of L-LTP 24 h after the tetanic stimulus protocol of Fig. 3, *i.e.*, the increase in the synaptic weight W. With standard parameter values, W = 0.127 prior to tetanus, and 0.303 24 h after tetanus. Thus, the control value of R is 0.176. Small (0.1%) changes in each parameter $p_i$ were then made to calculate the $S_i$s. The parameters were those in Table I as well as the basal (unstimulated) levels of [cAMP], [$Ca^{2+}_{syn}$], and [$Ca^{2+}_{nuc}$].

All of the $S_i$s were found to have an absolute value < 3. Thus, the model is not unduly sensitive to variations in any one parameter. The range of $S_i$s was (-2.30, 2.55). Of the 46 $S_i$s, 10 had an absolute value above 1. Eight $S_i$s had absolute value > 1.3, corresponding to the parameters [Raf]$_{tot}$ ($S_i$ = 2.55), $k_{f,MAPKK}$ ($S_i$ = 2.55), $k_{b,MAPKK}$ ($S_i$ = -2.30), $k_{b,Raf}$ ($S_i$ = -2.21), $k_{f,Raf}$(basal) ($S_i$ = 1.57), $k_{f,MAPK}$ ($S_i$ = 1.48), $K_{camp}$ ($S_i$ = -1.71) , and [cAMP]$_{basal}$ ($S_i$ = -1.48). All of these parameters except $K_{camp}$ and [cAMP]$_{basal}$ characterize the kinetics of the MAP kinase cascade. As discussed above, multiple phosphorylations within this cascade generate a supralinear relationship between Raf activation and MAPK activation. Thus, the magnitude of L-LTP induction exhibits a rather sensitive dependence on kinetic parameters of the MAPK cascade.

The relative sensitivities calculated with small parameter changes may not always predict the response of the model to larger parameter changes. Therefore, the calculation of the $S_i$s was repeated, using substantial (40%) increases in each parameter $p_i$. Interestingly, an overall damping of the $S_i$s was observed. Of the 46 $S_i$s, 41 decreased in absolute value. The $S_i$ range decreased to (-1.7, 0.53). Only four $S_i$s had absolute value > 1.0, corresponding to the parameters $k_{b,MAPKK}$ ($S_i$ = -1.69), $k_{b,Raf}$ ($S_i$ = -1.68), $k_{b,MAPK}$ ($S_i$ = -1.08), and $K_{camp}$ ($S_i$ = -1.35). The magnitude of L-LTP remains rather sensitive to MAPK cascade kinetics. The damping of the $S_i$s with larger parameter changes suggests the model is reasonably robust to parameter variability, as is necessary for a plausible model of intracellular signaling and responses to stimuli.

Can the model predict a pattern of tetanic inter-stimulus intervals (ISIs) that is optimal for induction of L-LTP? To examine this question, we first determined the dependence of L-LTP on the ISI for a group of three tetani, simulated as for Fig. 3, with the ISI varying from 0 to 300 min in steps of 1 min. For each simulation, the amount of L-LTP (the increase of W) was determined 24 hrs post-tetanus. Only a small enhancement of L-LTP by stimulus spacing was found. L-LTP was 138% for an ISI of 1 min, increasing slightly to a peak of 147% for ISI's of 9-15 min. Above 15 min L-LTP declined smoothly, to 100% for an ISI of 60 min and 36% for an ISI of 300 min. The model therefore predicts relatively little enhancement of hippocampal tetanic L-LTP when the ISI is increased from ~1 min to 5 min or longer.

However, the observed decline of L-LTP for long ISIs (≥60 min) suggests that for long ISIs, a strong enhancement of L-LTP can be produced by grouping stimuli into bursts. To explore this enhancement, we simulated six tetani, delivered in two protocols: 1) equal separation by ISIs of 3 hrs, *vs.* 2) two bursts of three tetani, with ISIs of 10 min within bursts and 860 min between bursts. Both protocols have a duration of 15 hrs. Twenty-four hrs after stimuli, the L-LTP induced by Protocol 1 was 95%, whereas Protocol 2 induced a much greater L-LTP, 250%. Similar enhancements of L-LTP (not shown) were observed for grouping of stimuli into four-tetanus bursts, and for replacement of tetanic stimuli by 10-min chemical stimuli. Two-tetanus bursts induce much less L-LTP as discussed previously, and bursts of more than four tetani induce little additional L-LTP due to depletion of the precursor protein P (Eq. 19). Therefore, the model predicts that a stimulus pattern maximizing induction of L-LTP can be obtained by grouping stimuli into bursts of 3-4 tetani each. Within each burst, the ISI should be 10-15 min.

## Simulations of L-LTP inhibition

Empirically, inhibition of CAMKII during and after stimuli blocks LTP induced by tetani (19) or by a pairing protocol (72). However, if the CAMKII inhibitor was perfused postsynaptically immediately <u>after</u> either stimulus protocol, no inhibition of LTP was observed. The model can simulate these observations. Figure 4 illustrates that a block of L-LTP results when CAMKII activity is inhibited for 1 hr during and after three tetani. In contrast, if the 1-hr CAMKII inhibition is assumed to begin 5 min after the tetani, L-LTP is not significantly attenuated. The window during which CAMKII activation is required is narrow, comprising the tetani and only a few minutes afterwards. Therefore, in the model, the rapid decay of CAMKII activity in ~ 5 min after tetanus (Fig. 3A) represents the disappearance of the <u>requirement</u> of CAMKII activity for L-LTP. Recent data suggest CAMKII activity may decay rapidly. Although hippocampal CAMKII <u>phosphorylation</u> persists for at least 30 min after tetani (73), the <u>activity</u> of CAMKII appears to decay within ~5 min after tetanic or chemical stimuli (74).

*Place Figure 4 near here*

Figure 4 also illustrates the effect on tetanic L-LTP of simulated inhibition of MAPK signaling by the commonly used compounds U0126 or PD98059, which block MAPKK activation (75). Strong attenuation is simulated (Fig. 4) if inhibition of MAPKK activation is modeled as a 90% reduction in the rate constant $k_{f, MAPKK}$ (Eqs. 7-8) during tetani and for 10 min after (as noted in Methods, such parameter alterations are applied identically to the somatic and synaptic MAPK cascades). Experimentally, inhibiting MAPKK activation during and after tetanic stimulation blocks L-LTP induction (30) Theta-burst L-LTP is also strongly attenuated by U0126 if this inhibitor is present during and for ~10 min after stimulus (14). In the model, the dual action of MAPK to phosphorylate a transcription factor (TF-2) and a synaptic substrate (Tag-3) is necessary for strong L-LTP attenuation. A model variant in which MAPK phosphorylates only one substrate retains considerable residual L-LTP (not shown).

Figure 4 also illustrates inhibition of L-LTP due to CAMKIV inhibition during and after tetanus. Empirically, transgenic mice expressing dominant-negative CAMKIV exhibit impaired L-LTP (24). In the simulation, CAMKIV was not inhibited prior to tetanus, although in the mice CAMKIV activity should be reduced at all times. In the model, inhibition of CAMKIV prior to tetanus reduces gene expression (the concentration of GPROD), thereby decreasing the basal value of the synaptic weight W, whereas experimentally, dominant negative CAMKIV does not reduce basal synaptic strength (25). This contradiction suggests that *in vivo*, a compensatory homeostatic mechanism preserves basal synaptic weights. For simplicity, the current model does not hypothesize a homeostatic mechanism. In the model, the lack of a homeostatic mechanism similarly leads to diminished basal synaptic strength with CAMKII, MAPK, or PKA inhibition. A planned extension will incorporate homeostatic regulation of basal synaptic strength, which may maintain neuronal activity and synaptic drive near set points (76).

Antisense *Arg3.1/Arc* mRNA oligonucleotides inhibited tetanic L-LTP by 40-60% (77). No effect was seen on baseline synaptic strength. To simulate this experiment, the rate of GPROD synthesis (Eq. 17) was decreased by 60% during and after three tetani. This alteration reduced the peak of [GPROD] by 59%, similar in magnitude to the empirical reduction in Arg3.1/Arc protein (77). Simulated L-LTP was reduced by 53%.

Tetanic L-LTP is blocked by a PKA inhibitor peptide, PKI (78). In the model, tetanic L-LTP was blocked when [PKA$_{act}$] was reduced by 90% during and after stimulation. Empirically, tetanic L-LTP was also blocked by a brief application of RpcAMP, which competitively inhibits cAMP's activation of PKA (68). RpcAMP was washed out after the tetanus. We attempted to simulate this experiment by terminating PKA inhibition five minutes after 3 simulated tetani. However, this did not block L-LTP. Five minutes after the tetani, phosphorylation of the CAMKII and MAPK synaptic tag substrates remained high. When PKA inhibition was terminated, the PKA substrate was significantly phosphorylated by basal PKA activity. The synaptic tag variable therefore increased, and overlapped with increased synthesis of GPROD, inducing L-LTP.

One possible explanation for the experimental block of L-LTP by brief RpcAMP applications is that RpcAMP inhibits PKA-independent activation of the MAPK signaling cascade. We therefore examined whether simulated L-LTP was inhibited if both PKA activity and MAPKK activation (k$_{f, MAPKK}$) were reduced by 90% during three tetani and for 5 min after. These reductions sufficed to inhibit L-LTP by 81%. There is experimental support for the suggestion that RpcAMP inhibits PKA-independent activation of MAPK. Activation by cAMP of the GTP-binding protein Rap1 can contribute to Raf activation (42) and this pathway appears independent of PKA (58,60).

## Simulation of theta-burst, pairing-induced, and chemical L-LTP

Figure 5A illustrates that the model simulates similar amounts of L-LTP for four stimulus protocols. L-LTP is taken to be the increase in W above baseline 24 h after each protocol. The largest potentiation (145%) is for tetanic L-LTP induction. A theta-burst stimulus (TBS) protocol was also simulated, yielding L-LTP of 89%, which is similar to experimental values (14). Inhibition of MAPKK activation (reduction of k$_{f, MAPKK}$ by 90%) during and for 10 min after TBS attenuated L-LTP by 81%. A similar attenuation was observed experimentally (14). We also simulated the L-LTP induction protocol used in (54), which pairs stimulation of two synapses. Substantial L-LTP (106%) resulted. The relatively weak electrical stimuli of the pairing protocol yield lower nuclear Ca$^{2+}$ and less CAMKIV activation. Therefore, to obtain substantial gene induction ([GPROD] elevation) and consequent L-LTP, the pairing protocol was assumed to strongly activate Raf and consequently MAPK (k$_{f,Raf}$ was elevated to 0.16 min$^{-1}$ for 6 min as described in Methods). The strong MAPK activation compensated for the weak CAMKIV activation, yielding substantial induction of GPROD and L-LTP. An experimental prediction follows. Pairing-induced L-LTP should be less inhibited than tetanic L-LTP after dominant negative CAMKIV is introduced as in (25).

Experimentally, chemical L-LTP (chem-LTP) is induced by forskolin or BDNF, without electrical stimulation. We first attempted to model chem-LTP by activation of Raf and PKA, without elevation of $Ca^{2+}$. However, significant L-LTP could not be simulated, because without some CAMKII activation, the level of synaptic tag remains very low, and without CAMKIV activation, the gene product level [GPROD] remains very low. We therefore assumed that synaptic and nuclear $Ca^{2+}$ were slightly elevated during the 30-min chemical application. Elevations of 60 nM for $[Ca^{2+}_{syn}]$ and $[Ca^{2+}_{nuc}]$ were assumed. Substantial chem-LTP (139%) was then simulated. Similar L-LTP magnitudes are observed experimentally (14,79). Figure 6A-B illustrates the simulation of chem-LTP. A large increase in the synaptic tag variable TAG is seen, partly due to very strong PKA activation and almost complete phosphorylation of the PKA tag substrate Tag-P2. The CAMKII activation that phosphorylates Tag-P1 and contributes to TAG elevation is small compared to that in electrical stimulus protocols (Fig. 6A, rise in CAMKII activity slightly above baseline).

Empirically, it is plausible that forskolin or BDNF application elevates $[Ca^{2+}]$. In GnRH neurons, increased cAMP augments $[Ca^{2+}]$ (63). BDNF application to cultured hippocampal neurons increased $[Ca^{2+}]$, apparently due to IP3-gated $Ca^{2+}$ release from intracellular stores (64).

Inhibition of MAPKK activation by U0126 or PD98059 suffices to block chem-LTP even when the inhibitor is washed out immediately after BDNF or forskolin application (13,14). The model simulates this behavior. If MAPKK activation is inhibited by 90% during and for 5 min after the chemical stimulus, L-LTP is strongly attenuated (the increase in W is reduced by 78%, Fig. 6B).

We examined whether simulated theta-burst, pairing-induced, and chemical L-LTP exhibited threshold behavior, *i.e.*, a supralinear increase in the amount of L-LTP *vs.* the stimulus duration. The threshold for tetanic L-LTP was discussed above. We reduced the duration of the theta-burst, pairing, and chemical protocols by 40%. L-LTP was reduced by greater percentages; 80% (theta-burst), 68% (pairing), and 67% (chemical). These greater percentage reductions illustrate that a supralinear increase of L-LTP with stimulus duration exists for all protocols, and this supralinearity is steepest for the theta-burst protocol and the tetanic protocol.

## Simulation of synaptic tagging

We examined whether the model could simulate the primary synaptic tagging experiment presented in (22) (their Fig. 1). In that experiment, one synapse, synapse A, was first given three tetani (100 Hz for 1 sec, interstimulus interval of 10 min), inducing L-LTP. Thirty-five min later, protein synthesis was halted by anisomycin. A second synapse, synapse B, was then given three tetani. One hour separated the first tetanus to synapse A and that to synapse B. Despite the presence of anisomycin, synapse B underwent L-LTP. This experiment has been interpreted (21,22) as supporting the hypothesis of synaptic tagging, with synapse B "tagged" by the second set of tetani. Synapse B can then "capture" the gene products that were previously synthesized as a consequence of the tetani to synapse A.

To model this experiment, the model of Fig. 1 was extended to represent two synapses, as described in Methods (Model Development). For synapse A, the first set of three tetani activated synaptic kinases, somatic and nuclear MAPK, and GPROD synthesis, yielding substantial L-LTP (traces for TAG-A, [GPROD], and W(tetanic), Fig. 7B). No L-LTP of synapse B resulted, because kinases at synapse B were not activated. To model the effect of anisomycin, synthesis of GPROD was halted 35 min after the tetani to synapse A. The second set of tetani, to synapse B only and with anisomycin, had no effect on synapse A. However, these tetani activated kinases at synapse B, setting the synaptic tag (trace for TAG-B, Fig. 7B). Substantial L-LTP of synapse B resulted (115% increase in W(tagged), Fig. 7B) because the TAG-B time course for synapse B overlapped the GPROD time course resulting from prior stimulation of synapse A. The TAG-B time course subsequently decays within 3 hrs, similarly to data (21,22).

*Place Figure 7 near here*

## DISCUSSION

### A model of L-LTP induction clarifies the roles of essential biochemical nonlinearities

We have constructed a model assigning experimentally supported roles to kinases essential for the induction and expression of L-LTP. The model is useful to: 1) clarify the significance of the biochemical nonlinearities that are essential for amplifying a brief stimulus (elevation of $[Ca^{2+}]$) into a long-lasting increase in synaptic strength; 2) provide a framework for interpreting the effects of manipulations affecting L-LTP, such as kinase inhibition; and 3) predict outcomes of experiments to delineate mechanisms of L-LTP induction and expression.

In the model, L-LTP inducing stimuli are represented by separate increases in $[Ca^{2+}]$, [cAMP], and synaptic and somatic Raf activation. However, cAMP elevation in electrically stimulated neurons appears to follow $[Ca^{2+}]$ elevation and activation of adenylyl cyclase 1 and 8 (37,38), and Raf activation appears at least partly driven by $[Ca^{2+}]$ elevation (57). Therefore, the increase in synaptic weight seen in L-LTP is predominantly driven by very brief (~1-5 sec) increases in intracellular $[Ca^{2+}]$. As discussed in Results, the model represents a supralinear relationship between the stimulus of $Ca^{2+}$ elevation and the response of synaptic weight increase, and this supralinearity is essential for amplifying a brief $[Ca^{2+}]$ increase into a long-lasting increase in the synaptic weight W. The supralinearity also results in threshold dynamics, in that the amount of L-LTP increases steeply with the number of stimuli (see Results).

Empirically, a similar supralinear relationship between $[Ca^{2+}]$ elevation and synaptic weight increase has been found. Moderate stimuli, such as low-frequency electrical pulses, produce LTD, whereas with stronger stimuli, there is a crossover to LTP. The kinetic profiles of $Ca^{2+}$ signals generated by stimuli in cortical slices have recently been compared with the plasticity outcome (80). An abrupt crossover from LTD to LTP occurred when peak $[Ca^{2+}]$ increased over a relatively narrow range, ~ 0.7 to 1.0 μM. Such an abrupt crossover requires a supralinear correlation between peak $[Ca^{2+}]$ and LTP. Our model suggests that the convergence of multiple kinases, including CaM kinases, is important for this nonlinearity. The dependence of L-LTP on multiple kinases may also allow numerous physiological regulatory points for this fundamental process.

Sensitivity analysis indicated the model dynamics are not overly sensitive to variations in any parameter. The L-LTP induced by simulated tetani is most sensitive to kinetic parameters in the MAPK cascade. It is plausible that these parameters could serve as physiological control points regulating L-LTP induction. Altering the intracellular distribution of Raf or MAPKK, or their interactions with other proteins, could alter the available amounts of these enzymes ($[Raf]_{tot}$, $[MAPKK]_{tot}$) or their catalytic efficiencies ($k_{f,MAPK}$, $k_{f,MAPKK}$, $k_{f,Raf}$). There is significant interest in developing pharmacological agents to enhance memory formation (81,82,83). Simulations illustrated that the sensitivities of L-LTP to alterations in specific parameters such as the dephosphorylation rate constants for MAPKK and Raf ($k_{b,MAPKK}$, $k_{b,Raf}$) are substantial. A pharmacological agent that inhibits the dephosphorylation and deactivation of MAPKK or Raf might significantly enhance L-LTP induction and the formation of LTM.

The model predicts that maximal induction of L-LTP can be achieved by grouping stimuli into bursts. For tetanic stimuli or brief chemical applications, bursts of 3-4 stimuli, with an inter-stimulus interval of 10-15 min, are optimal for simulated L-LTP induction. There is little experimental data to compare this prediction to. One study reports that for a burst of 3 tetani, an inter-stimulus interval of ~10 min is indeed optimal for LTP (84). However, these authors restricted their assay to E-LTP (45 min post-stimulus).

Following L-LTP, W decays very slowly towards basal values, at a rate governed by the large time constant for decay of W. $\tau_W$ =140,000 min (3.2 months). Indeed, L-LTP can persist for months (2) although *in vivo*, depression due to competitive potentiation of other synapses (85), or homeostatic regulation of synaptic weights (76), might commonly eliminate L-LTP more rapidly.

Given that lifetimes of synaptic proteins *in vivo* are on the order of hours to days (86), maintenance of L-LTP for weeks or months must rely on processes that can compensate for molecular turnover. These processes are not yet well characterized, and are not currently represented in the model. Bistable molecular synaptic switches have been proposed that, if set to an active state, might retain this state and maintain high synaptic strength for months or longer. Three proposed switches are: 1) A positive feedback loop based on mutually reinforcing activation of MAPK, protein kinase C, and phospholipase A2 (5, 87); 2) A switch whereby transient activation of PKA phosphorylates a critical number of AMPA receptors, sufficient to saturate phosphatase activity, so that basal PKA activity can then maintain phosphorylation of these receptors (88); 3) A positive feedback loop in which transient enhancement of translation of the elongation factor eEF1A leads to self-reinforcing, enhanced translation of eEF1A and other mRNAs necessary for L-LTP (89,90). The kinase activation events and gene induction represented in our model could serve as input to these proposed switches. Transient MAPK and PKA activation could respectively activate switches 1 and 2. Switch 3 could be activated following MAPK activation, because neuronal MAPK activation leads to phosphorylation of multiple translation factors (16), plausibly enhancing translation of eEF1A. An alternative proposal for long-term maintenance of L-LTP posits recurrent activations of the synaptic networks that store memories, perhaps during sleep (91,92). These episodes of activity could drive repeated L-LTP events that maintain synaptic strengths.

**The model represents key signaling pathways involved in L-LTP induction**

We believe that the model represents the most commonly proposed roles of kinases essential for L-LTP, in particular CAMKII, CAMKIV, PKA, and ERK isoform(s) of MAPK. However, these representations are qualitative and do not consider many details of kinase regulation or function. For example, MAPK is represented as directly phosphorylating a nuclear transcription factor, not considering activation by MAPK of the CREB kinases RSK-2 or mitogen and stress-activated kinase (MSK) (93,94). Nevertheless, we believe our representations of kinases suffice to illustrate important dynamic elements, such as supralinear stimulus-response relationships that appear essential for L-LTP induction. Also, the model can simulate a variety of kinase inhibitor experiments. Simulations do fail to account for the preservation of basal synaptic strength in the presence of dominant negative CAMKIV or of inhibitors of the other kinases. However, these failures are useful, indicating that a more comprehensive model will need to incorporate homeostatic mechanisms for preserving basal synaptic strengths.

The model does not represent all significant biochemical pathways involved in L-LTP. One such pathway is activation of PKM/PKCζ due to translation of a PKM-specific mRNA (95) and consequent activation of p70s6 kinase (96), which upregulates dendritic translation near activated synapses (97). Another such pathway may be transport of phosphorylated CREB from dendrites to nucleus following synaptic stimulation (98). We believe, however, that the model provides a flexible framework to incorporate additional pathways as they are characterized. Very recently, phosphatidylinositol-3-kinase (PI3K) inhibition has been reported to reverse L-LTP (99). It may, therefore, be useful to incorporate a representation of the PI3K signaling pathway.

**The model suggests experimental predictions**

Experimental predictions could either falsify or support key assumptions of our model, as follows: 1) The model assumes that forskolin or BDNF exposure significantly elevates $Ca^{2+}$, activating CaM kinases necessary for L-LTP (Fig. 6). As discussed in Results, forskolin or BDNF have been observed to elevate neuronal $Ca^{2+}$. However, those experiments did not involve L-LTP induction protocols. If $Ca^{2+}$ elevation is necessary for chem-LTP, inhibition of CAMKII (72), or introduction of dominant negative CAMKIV (25), should impair chem-LTP. Fluorescent $Ca^{2+}$ indicators should also reveal a significant, but modest, increase in $[Ca^{2+}]$. 2) To simulate a block of L-LTP due to a brief application of the inactive cAMP analogue RpcAMP, the model assumes RpcAMP inhibits activation of the MAPK cascade. If this assumption is correct, RpcAMP should inhibit experimental activation of ERK I/II isoforms of MAPK due to tetanic stimulation. As noted previously, there is evidence that cAMP can activate the neuronal MAPK cascade (58,42).

Additional predictions may help to clarify the role of PKA in L-LTP. Although PKA activity has been reported necessary for setting a synaptic tag (20), the PKA inhibitor used, KT5720, is not very selective. It inhibits a number of kinases, including MAP kinase kinase 1, at least as strongly as PKA (75). The peptide inhibitor of PKA, PKI, has been infused into postsynaptic pyramidal neurons during L-LTP recordings (78). A synaptic tagging experiment similar to Fig. 1D of (22), with anisomycin present during the second set of tetani to synapse B, might be repeated with PKI infused prior to tetanizing synapse B. The model assumes PKA activity is necessary for tagging, and predicts that L-LTP of synapse B would be blocked by PKI. The model also assumes that PKA activity is necessary for nuclear translocation of MAPK. If this is correct, infusion of PKI should block MAPK translocation observed following theta-burst stimulation (14) or BDNF application (29). Infusion of PKI should also block chem-LTP.

The model appears helpful in identifying similarities, and at least one major difference, between mechanisms of L-LTP induction and another form of long-lasting synaptic strengthening, long-term facilitation (LTF) of synapses in the mollusk *Aplysia* and other invertebrates (100). In *Aplysia*, LTF of synapses from sensory to motor neurons is induced by spaced applications of serotonin (5-HT). Typically five 5-min pulses of 5-HT are used (101,102). As with L-LTP, activation and nuclear translocation of an ERK isoform of MAPK appear necessary for LTF (102,103). LTF exhibits a "threshold" nonlinearity in that five pulses of 5-HT induce LTF, but four do not (104). As discussed for L-LTP (see Results), such threshold behavior, as well as the ability of brief inputs (5-HT applications) to produce long-lasting synaptic change, suggest supralinear stimulus-response relationships. The requirement for multiple phosphorylations to activate MAPKK and MAPK may generate such a nonlinearity for *Aplysia* LTF and for L-LTP.

A major difference between LTF and L-LTP is that LTF induced by spaced 5-HT applications has not been found to require elevation of cytoplasmic or nuclear $Ca^{2+}$. Nevertheless, as with L-LTP, there is likely to be a supralinear convergence of activation of multiple kinases to induce LTF. PKA is activated during LTF induction (101), and can phosphorylate the CREB1 transcriptional activator. *Aplysia* ERK affects gene induction by phosphorylation of at least one transcription factor, CREB2 (102). As with L-LTP, PKA activity appears necessary to set a synaptic tag allowing LTF (105,106). Experiments like those suggested for L-LTP could also clarify the role of PKA in LTF. Injection of PKI into *Aplysia* sensory neurons would be predicted to block synaptic tagging and nuclear ERK translocation.

## REFERENCES

1. Lynch, M. A. 2004. Long-term potentiation and memory. *Physiol. Rev.* 84:87-136.

2. Abraham, W. C., B. Logan, J. M. Greenwood, and M. Dragunow. 2002. Induction and experience-dependent consolidation of stable long-term potentiation lasting months in the hippocampus. *J. Neurosci.* 22:9626-9634.

3. Kikuchi, S., K. Fujimoto, N. Kitagawa, T. Fuchikawa, M. Abe, K. Oka, K. Takei, and M. Tomita. 2003. Kinetic simulation of signal transduction system in hippocampal long-term potentiation with dynamic modeling of protein phosphatase 2A. *Neural Netw.* 16:1389-1398.

4. Lisman, J. E., and A. M. Zhabotinsky. 2001. A model of synaptic memory: a CaMKII/PP1 switch that potentiates transmission by organizing an AMPA receptor anchoring assembly. *Neuron* 31:191-201.

5. Bhalla, U. S., and R. Iyengar. 1999. Emergent properties of networks of biological signaling pathways. *Science* 283:381-387.

6. Nguyen, P. G., T. Abel, and E. R. Kandel. 1994. Requirement of a critical period of transcription for induction of a late phase of LTP. *Science* 265:1104-1107.

7. Bozon, B., A. Kelly, S. A. Josselyn, A. J. Silva, S. Davis, and S. Laroche. 2003. MAPK, CREB and zif268 are all required for the consolidation of recognition memory. *Philos. Trans. R. Soc. Lond. B Biol. Sci.* 358:805-814.

8. Hevroni, D., A. Rattner, M. Bundman, D. Lederfein, A. Gabarah, M. Mangelus, M. A. Silverman, H. Kedar, C. Naor, M. Komuc, T. Hanoch, R. Seger, L. E. Theill, E. Nedivi, G. Richter-Levin, and Y. Citri. 1998. Hippocampal plasticity involves extensive gene induction and multiple cellular mechanisms. *J. Mol. Neurosci.* 10:75-98.

9. Auer, R. N., M. L. Jensen, and I. Q. Whishaw. 1989. Neurobehavioral deficit due to ischemic brain damage limited to half of the CA1 sector of the hippocampus. *J. Neurosci.* 9:1641-1647.

10. Zola-Morgan, S., L. R. Squire, and D. G. Amaral. 1986. Human amnesia and the medial temporal region: enduring memory impairment following a bilateral lesion limited to field CA1 of the hippocampus. *J. Neurosci.* 6:2950-2967.

11. Tsien, J. Z., P. T. Huerta, and S. Tonegawa. 1996. The essential role of hippocampal CA1 NMDA receptor-dependent synaptic plasticity in spatial memory. *Cell* 87:1327-1338.

12. Raser, J. M., and E. K. O'Shea. 2005. Noise in gene expression: origins, consequences, and control. *Science* 309:2010-2013.

13. Ying, S. W., M. Futter, K. Rosenblum, M. J. Webber, S. P. Hunt, T. V. Bliss, and C. R. Bramham. 2002. Brain-derived neurotrophic factor induces long-term potentiation in intact adult hippocampus: requirement for ERK activation coupled to CREB and upregulation of Arc synthesis. *J. Neurosci.* 22:1532-1540.

14. Patterson, S. L., C. Pittenger, A. Morozov, K. C. Martin, H. Scanlin, C. Drake, and E. R. Kandel. 2001. Some forms of cAMP-mediated long-lasting potentiation are associated with release of BDNF and nuclear translocation of phospho-MAP kinase. *Neuron* 32:123-140.

15. Atkins, C. M., N. Nozaki, Y. Shigeri, and T. R. Soderling. 2004. Cytoplasmic polyadenylation element binding protein-dependent protein synthesis is regulated by calcium/calmodulin-dependent protein kinase II. *J. Neurosci.* 24:5193-5201.

16. Kelleher, R. J., A. Govindarajan, H. Y. Jung, H. Kang, and S. Tomegawa. 2004. Translational control by MAPK signaling in long-term synaptic plasticity and memory. *Cell* 116:467-479.

17. Bradshaw, K. D., N. J. Emptage, and T. V. Bliss. 2003a. A role for dendritic protein synthesis in hippocampal late LTP. *Eur. J. Neurosci.* 18:3150-3152.

18. Cammalleri, M., R. Lutjens, F. Berton, A. R. King, C. Simpson, W. Francesconi, and P. P. Sanna. 2003. Time-restricted role for dendritic activation of the mTOR-p70S6K pathway in the induction of late-phase long-term potentiation in the CA1. *Proc. Natl. Acad. Sci. USA* 100:14368-14373.

19. Chen, H. X., N. Otmakhov, S. Strack, R. J. Colbran, and J. E. Lisman. 2001. Is persistent activity of calcium/calmodulin-dependent kinase required for the maintenance of LTP? *J. Neurophysiol.* 85:1368-1376.

20. Barco, A., J. M. Alarcon, and E. R. Kandel. 2002. Expression of constitutively active CREB facilitates the late phase of long-term potentiation by enhancing synaptic capture. *Cell* 108:689-703.

21. Frey, U., and R. G. Morris. 1998. Weak before strong, dissociating synaptic tagging and plasticity-factor accounts of late-LTP. *Neuropharmacology* 37:545-552.

22. Frey, U., and R. G. Morris. 1997. Synaptic tagging and long-term potentiation. *Natur*e 385:533-536.

23. Wu, G. Y., K. Deisseroth, and R. W. Tsien. 2001. Activity-dependent CREB phosphorylation: convergence of a fast, sensitive calmodulin kinase pathway and a slow, less sensitive mitogen-activated protein kinase pathway. *Proc. Natl. Acad. Sci. USA* 98:2808-2813.

24. Impey, S., A. L. Fong, Y. Wang, J. R. Cardinaux, D. M. Fass, K. Obrietan, G. A. Wayman, D. R. Storm, T. R. Soderling, and R. H. Goodman. 2002. Phosphorylation of CBP mediates transcriptional activation by neural activity and CaM kinase IV. *Neuron* 34:235-244.

25. Kang, H., L. D. Sun, C. M. Atkins, T. R. Soderling, M. A. Wilson, and S. Tonegawa. 2001. An important role of neural activity-dependent CaMKIV signaling in the consolidation of long-term memory. *Cell* 106:771-783.

26. Davis, S., P. Vanhoutte, C. Pages, J. Caboche, and S. Laroche. 2000. The MAPK/ERK cascade targets both Elk-1 and cAMP response element-binding protein to control long-term potentiation-dependent gene expression in the dentate gyrus *in vivo*. *J. Neurosci.* 20:4563-4572.

27. Jones, M. W., M. L. Errington, P. J. French, A. Fine, T. V. Bliss, S. Garel, P. Chamay, B. Bozon, S. Laroche, and S. Davis. 2001. A requirement for the immediate early gene Zif268 in the expression of late LTP and long-term memories. *Nat. Neurosci.* 4:289-296.

28. Waltereit, R., B. Dammermann, P. Wulff, J. Scafidi, U. Staubli, G. Kauselmann, M. Bundman, and D. Kuhl. 2001. *Arg3.1/Arc* mRNA induction by $Ca^{2+}$ and cAMP requires protein kinase A and mitogen-activated protein kinase/extracellular regulated kinase activation. *J. Neurosci.* 21:5484-5493.

29.   Rosenblum, K., M. Futter, K. Voss, M. Erent, P. A. Skehel, P. French, L. Obosi, M. W. Jones, and T. V. Bliss. 2002. The role of extracellular regulated kinases I/II in late-phase long-term potentiation. *J. Neurosci.* 22:5432-5441.

30.   English, J. D., and J. D. Sweatt. 1997. A requirement for the mitogen-activated protein kinase cascade in hippocampal long term potentiation. *J. Biol. Chem.* 272:19103-19106.

31.   Impey, S., K. Obrietan, S. T. Wong, S. Poser, S. Yano, G. Wayman, J. C. Deloulme, G. Chan, and D. R. Storm. 1998. Cross talk between ERK and PKA is required for $Ca^{2+}$ stimulation of CREB-dependent transcription and ERK nuclear translocation. *Neuron* 21:869-883.

32.   Chetkovich, D. M., R. Gray, D. Johnston, and J. D. Sweatt. 1991. N-methyl-D-aspartate receptor activation increases cAMP levels and voltage-gated $Ca^{2+}$ channel activity in area CA1 of hippocampus. *Proc. Natl. Acad. Sci. USA* 88:6467-6471.

33.   Roberson, E. D., and J. D. Sweatt. 1996. Transient activation of cyclic AMP-dependent protein kinase during hippocampal long-term potentiation. *J. Biol. Chem.* 271:30436-30441.

34.   Pokorska, A., P. Vanhoutte, F. J. Arnold, F. Silvagno, G. E. Hardingham, and H. Bading. 2003. Synaptic activity induces signalling to CREB without increasing global levels of cAMP in hippocampal neurons. *J. Neurochem.* 84:447-452.

35.   Nguyen, P. G., and E. R. Kandel. 1996. A macromolecular synthesis-dependent late phase of long-term potentiation requiring cAMP in the medial perforant pathway of rat hippocampal slices. *J. Neurosci.* 16:3189-3198.

36.   Frey, U., Y. Y. Huang, and E. R. Kandel. 1993. Effects of cAMP simulate a late stage of LTP in hippocampal CA1 neurons. *Science* 260:1661-1664.

37.   Wang, H., V. V. Pineda, G. C. Chan, S. T. Wong, L. J. Muglia, and D. R. Storm. 2003. Type 8 adenylyl cyclase is targeted to excitatory synapses and required for mossy fiber long-term potentiation. *J Neurosci* 23:9710–9718.

38.   Wong, S. T., J. Athos, X. A. Figueroa, V. V. Pineda, M. L. Schaefer, C. C. Chavkin, L. J. Muglia, and D. R. Storm. 1999. Calcium-stimulated adenylyl cyclase activity is critical for hippocampus-dependent long-term memory and late phase LTP. *Neuron* 23:787–798.

39.   Bradshaw, J. M., Y. Kubota, T. Meyer, and H. Schulman. 2003b. An ultrasensitive $Ca^{2+}$/calmodulin-dependent protein kinase II-protein phosphatase 1 switch facilitates specificity in postsynaptic calcium signaling. *Proc. Natl. Acad. Sci. USA* 100:10512-10517.

40.   Herberg, F. W., S. S. Taylor, and W. R. Dostmann. 1996. Active site mutations define the pathway for the cooperative activation of cAMP-dependent protein kinase. *Biochemistry* 35:2934-2942.

41.   Agell, N., O. Bachs, N. Rocamora, and P. Villalonga. 2002. Modulation of the Ras/Raf/MEK/ERK pathway by $Ca^{2+}$ and calmodulin. *Cell. Signalling* 14:649-654.

42. Dugan, L. L., J. S. Kim, Y. Zhang, R. D. Bart, Y. Sun, D. M. Holtzman, and D. H. Gutmann. 1999. Differential effects of cAMP in neurons and astrocytes. Role of B-raf. *J. Biol. Chem.* 274:25842-25848.

43. Pettigrew, D. B., P. Smolen, D. A. Baxter, and J. H. Byrne. 2005. Dynamic properties of regulatory motifs associated with induction of three temporal domains of memory in *Aplysia*. *J. Comput. Neurosci.* 18:163-181.

44. Komiyama, N. H., A. M. Watabe, H. J. Carlisle, K. Porter, P. Charlesworth, J. Monti, D. J. Strathdee, C. M. O'Carroll, S. J. Martin, R. G. Morris, T. J. O'Dell, and S. G. Grant. 2002. SynGAP regulates ERK/MAPK signaling, synaptic plasticity, and learning in the complex with postsynaptic density 95 and NMDA receptor. *J. Neurosci.* 22:9721-9732.

45. Banko, J. L., L. Hou, and E. Klann. 2004. NMDA receptor activation results in PKA- and ERK-dependent Mnk1 activation and increased eIF4E phosphorylation in hippocampal area CA1. *J. Neurochem.* 91:462-470.

46. Atkins, C. M., M. A. Davare, M. C. Oh, V. Derkach, and T. R. Soderling. 2005. Bidirectional regulation of cytoplasmic polyadenylation element-binding protein phosphorylation by $Ca^{2+}$/calmodulin – dependent protein kinase II and protein phosphatase I during hippocampal long-term potentiation. *J. Neurosci.* 25:5604-5610.

47. Sanna, P. P., M. Cammalleri, F. Berton, C. Simpson, R. Lutiens, F. E. Bloom, and W. Francesconi. 2002. Phosphatidylinositol 3-kinase is required for the expression but not for the induction or the maintenance of long-term potentiation in the hippocampal CA1 region. *J. Neurosci.* 22:3359-3365.

48. Kanzaki, M., S. Mora, J. B. Hwang, A. R. Saltiel, and J. E. Pessin. 2004. Atypical protein kinase C (PKCzeta/lambda) is a convergent downstream target of the insulin-stimulated phosphatidylinositol 3-kinase and TC10 signaling pathways. *J. Cell Biol.* 164:279-290.

49. Le Good, J. A., W. H. Ziegler, D. B. Parekh, D. R. Alessi, P. Cohen, and P. J. Parker. 1998. Protein kinase C isotypes controlled by phosphoinositide 3-kinase through the protein kinase PDK1. *Science* 281:2042-2045.

50. Doskeland, S. O., and D. Ogreid. 1981. Binding proteins for cyclic AMP in mammalian tissue. *Int. J. Biochem.* 13:1-19.

51. Malenka, R. C., B. Lancaster, and R. S. Zucker. 1992. Temporal limits on the rise in postsynaptic calcium required for the induction of long-term potentiation. *Neuron* 9:121-128.

52. Pologruto, T. A., R. Yasuda, and K. Svoboda. 2004. Monitoring neural activity and $[Ca^{2+}]$ with genetically encoded $Ca^{2+}$ indicators. *J. Neurosci.* 24:9572-9579.

53. Hyrc, K., Z. Rzeszotnik, B. R. Kennedy, and M. P. Goldberg. 2005. Why do low and high affinity calcium indicators report cytosolic $Ca^{2+}$ concentrations differently in NMDA but not in AMPA – stimulated neurons? *Soc. Neurosci. Abstr.* 467.5.

54. Huang, Y. Y., C. Pittenger, and E. R. Kandel. 2004. A form of long-lasting, learning-related synaptic plasticity in the hippocampus induced by heterosynaptic low-frequency pairing. *Proc. Natl. Acad. Sci. USA* 101:859-864.

55. Vincent, P., and D. Brusciano. 2001. Cyclic AMP imaging in neurons in brain slice preparation. *J. Neurosci. Meth.* 108:189-198.

56. Bacskai, B. J., B. Hochner, M. Mahaut-Smith, S. R. Adams, B. Kaang, E. R. Kandel, and R. Y. Tsien. 1993. Spatially resolved dynamics of cAMP and protein kinase A subunits in *Aplysia* sensory neurons. *Science* 260:222-226.

57. Schmitt, J. M., E. S. Guire, T. Saneyoshi, and T. R. Soderling. 2005. Calmodulin-dependent kinase kinase/calmodulin kinase I activity gates extracellular-regulated kinase-dependent long-term potentiation. *J. Neurosci.* 25:1281-1290.

58. Morozov, A., I. A. Muzzio, R. Bourtchouladze, N. Van-Strien, K. Lapidus, D. Yin, D. G. Winder, J. P. Adams, and J. D. Sweatt. 2003. Rap1 couples cAMP signaling to a distinct pool of p42/44MAPK regulating excitability, synaptic plasticity, learning, and memory. *Neuron* 39:309-325.

59. Grewal, S. S., A. M. Horgan, R. D. York, G. S. Withers, G. A. Banker, and P. J. Stork. 2000. Neuronal calcium activates a Rap1 and B-Raf signaling pathway via the cyclic adenosine monophosphate-dependent protein kinase. *J. Biol. Chem.* 275:3722-3728.

60. Kawasaki, H., G. M. Springett, N. Mochizuki, S. Toki, M. Nakaya, M. Matsuda, D. E. Housman, and A. M. Graybiel. 1998. A family of cAMP-binding proteins that directly activate Rap1. *Science* 282:2275-2279.

61. Yang, L., L. Mao, Q. Tang, S. Samdani, Z. Liu, and J. Q. Wang. 2004. A novel $Ca^{2+}$-independent pathway to extracellular signal-regulated protein kinase by coactivation of NMDA receptors and metabotropic glutamate receptor 5 in neurons. *J. Neurosci.* 24:10846-10857.

62. Iida, N., K. Namikawa, H. Kiyama, H. Ueno, S. Nakamura, and S. Hattori. 2001. Requirement of Ras for the activation of mitogen-activated protein kinase by calcium influx, cAMP, and neurotrophin in hippocampal neurons. *J. Neurosci.* 21:6459-6466.

63. Kaneishi, K., Y. Sakuma, H. Kobayashi, and M. Kato. 2002. 3',5'-cyclic adenosine monophosphate augments intracellular $Ca^{2+}$ concentration and gonadotropin-releasing hormone (GnRH) release in immortalized GnRH neurons in an $Na^+$-dependent manner. *Endocrinology* 143:4210-4217.

64. Li, Y. X., Y. Zhang, H. A. Lester, E. M. Schuman, and N. Davidson. 1998. Enhancement of neurotransmitter release induced by brain-derived neurotrophic factor in cultured hippocampal neurons. *J. Neurosci.* 18:10231-10240.

65. Press, W. H., S. A. Teukolsky, W. T. Vetterling, and B. P. Flannery. 1992. Numerical Recipes in Fortran 77. Cambridge University Press, Cambridge.

66. Ahmed, T., and J. U. Frey. 2005. Plasticity-specific phosphorylation of CAMKII, MAP-kinases, and CREB during late-LTP in rat hippocampal slices *in vitro. Neuropharmacology*, in press.

67. Liu, J., K. Fukunaga, H. Yamamoto, K. Nishi, and E. Miyamoto. 1999. Differential roles of $Ca^{2+}$/calmodulin-dependent protein kinase II and mitogen-activated protein kinase activation in hippocampal long-term potentiation. *J. Neurosci.* 19:8292-8299.

68. Woo, N. H., S. N. Duffy, T. Abel, and P. V. Nguyen. 2000. Genetic and pharmacological demonstration of differential recruitment of cAMP-dependent protein kinases by synaptic activity. *J. Neurophysiol.* 84:2739-2745.

69. Huang, C. Y., and J. E. Ferrell. 1996. Ultrasensitivity in the mitogen-activated protein kinase cascade. *Proc. Natl. Acad. Sci. USA* 93:10078-10083.

70. Frank, P. M. 1978. Introduction to System Sensitivity Theory. pp. 9-10. Academic Press, New York.

71. Beck, J. V., and K. J. Arnold. 1977. Parameter Estimation in Engineering and Science. pp. 17-24, 481-487. John Wiley, New York.

72. Otmakhov, N., L. C. Griffith, and J. E. Lisman. 1997. Postsynaptic inhibitors of calcium/calmodulin-dependent protein kinase type II block induction but not maintenance of pairing-induced long-term potentiation. *J. Neurosci.* 17:5357-5365.

73. Ouyang, Y., D. Kantor, K. M. Harris, E. M. Schuman, and M. B. Kennedy. 1997. Visualization of the distribution of autophosphorylated calcium/calmodulin-dependent protein kinase II after tetanic stimulation in the CA1 area of the hippocampus. *J. Neurosci.* 17:5416-5427.

74. Lengyel, I., K. Voss, M. Cammarota, K. Bradshaw, V. Brent, K. P. Murphy, K. P. Giese, J. P. Rostas, and T. V. P. Bliss. 2004. Autonomous activity of CAMKII is only transiently increased following the induction of long-term potentation in the rat hippocampus. *Eur. J. Neurosci.* 20:3063-3072.

75. Davies, S. P., H. Reddy, M. Caivano, and P. Cohen. 2000. Specificity and mechanism of action of some commonly used protein kinase inhibitors. *Biochem. J.* 351:95-105.

76. Burrone, J., and V. N. Murthy. 2003. Synaptic gain control and homeostasis. *Curr. Opin. Neurobiol.* 13:560-567.

77. Guzowski, J. F., G. L. Lyford, G. D. Stevenson, F. P. Houston, J. L. McGaugh, P. F. Worley, and C. A. Barnes. 2000. Inhibition of activity-dependent arc protein expression in the rat hippocampus impairs the maintenance of long-term potentiation and the consolidation of long-term memory. *J. Neurosci.* 20:3993-4001.

78. Duffy, S. N., and P. V. Nguyen. 2003. Postsynaptic application of a peptide inhibitor of cAMP-dependent protein kinase blocks expression of long-lasting synaptic potentiation in hippocampal neurons. *J. Neurosci.* 23:1142-1150.

79. Wu, S. P., K. T. Lu, W. C. Chang, and P. W. Gean. 1999. Involvement of mitogen-activated protein kinase in hippocampal long-term potentiation. *J. Biomed. Sci.* 6:409-417.

80. Ismailov, I., D. Kalikulov, T. Inoue, and M. J. Friedlander. 2004. The kinetic profile of intracellular calcium predicts long-term potentiation and long-term depression. *J. Neurosci.* 24:9847-9861.

81. Gong, B., O. V. Vitolo, F. Trinchese, S. Liu, M. Shelanski, and O. Arancio. 2004. Persistent improvement in synaptic and cognitive functions in an Alzheimer mouse model after rolipram treatment. *J. Clin. Invest.* 114:1624-1634.

82. Zhang, H. T., Y. Zhao, Y. Huang, N. R. Dorairaj, L. J. Chandler, and J. M. O'Donnell. 2004. Inhibition of the phosphodiesterase 4 (PDE4) enzyme reverses memory deficits produced by infusion of the MEK inhibitor U0126 into the CA1 subregion of the rat hippocampus. *Neuropsychopharmacology* 29:1432-1439.

83. Barad, M., R. Bourtchouladze, D. G. Winder, H. Golan, and E. R. Kandel. 1998. Rolipram, a type IV-specific phosphodiesterase inhibitor, facilitates the establishment of long-lasting long-term potentiation and improves memory. *Proc. Natl. Acad. Sci. USA* 95:15020-15025.

84. Ajay, S. M., and U. S. Bhalla. 2004. A role for ERKII in synaptic pattern selectivity on the time-scale of minutes. *Eur. J. Neurosci.* 20:2671-2680.

85. Fonseca, R., U. V. Nagerl, R. G. Morris, and T. Bonhoeffer. 2004. Competing for memory: hippocampal LTP under regimes of reduced protein synthesis. *Neuron* 44:1011-1020.

86. Ehlers, M. D. 2003. Activity level controls postsynaptic composition and signaling via the ubiquitin-proteasome system. *Nat. Neurosci.* 6:231-242.

87. Bhalla, U. S. 2002. Mechanisms for temporal tuning and filtering by postsynaptic signaling pathways. *Biophys. J.* 83:740-752.

88. Hayer, A., and U. S. Bhalla. 2005. Molecular switches at the synapse emerge from receptor and kinase traffic. *PLOS Comput. Biol.* 1:137-154.

89. Blitzer, R. D., R. Iyengar, and E. M. Landau. 2005. Postsynaptic signaling networks: cellular cogwheels underlying long-term plasticity. *Biol. Psychiatry* 57:113-119.

90. Tsokas, P., E. A. Grace, P. Chan, T. Ma, S. Sealfon, R. Iyengar, E. Landau, and R. D. Blitzer. 2005. Local protein synthesis mediates a rapid increase in dendritic elongation factor 1A after induction of late long-term potentiation. *J. Neurosci.* 25:5833-5843.

91. Horn, D., N. Levy, and E. Ruppin. 1998. Memory maintenance via neuronal regulation. *Neural Comput.* 10:1-18.

92. Wittenberg, G. W., M. R. Sullivan, and J. Z. Tsien. 2002. Synaptic reentry reinforcement based network model for long-term memory consolidation. *Hippocampus* 12:637-642.

93. Rakhit, S., C. J. Clark, C. T. O'Shaughnessy, and B. J. Morris. 2005. NMDA and BDNF induce distinct profiles of extracellular regulated kinase (ERK), mitogen and stress activated kinase (MSK) and ribosomal S6 kinase (RSK) phosphorylation in cortical neurones. *Mol. Pharmacol.* 67:1158-1165.

94. Frodin, M., and S. Gammeltoft. 1999. Role and regulation of 90 kDa ribosomal S6 kinase (RSK) in signal transduction. *Mol. Cell. Endocrinol.* 151:65-77.

95. Hernandez, A. I., N. Blace, J. F. Crary, P. A. Serrano, M. Leitges, J. M. Libien, G. Weinstein, A. Tcherapov, and T. C. Sacktor. 2003. Protein kinase M zeta synthesis from a brain mRNA encoding an independent protein kinase C zeta catalytic domain. Implications for the molecular mechanism of memory. *J. Biol. Chem.* 238:40305-40316.

96. Romanelli, A., K. A. Martin, A. Toker, and J. Blenis. 1999. p70 S6 kinase is regulated by protein kinase Czeta and participates in a phosphoinositide 3-kinase-regulated signalling complex. *Mol. Cell. Biol.* 19:2921-2928.

97. Takei, N., N. Inamura, M. Kawamura, H. Namba, K. Hara, K. Yonezawa, and H. Nawa. 2004. Brain-derived neurotrophic factor induces mammalian target of rapamycin-dependent local activation of translation machinery and protein synthesis in neuronal dendrites. *J. Neurosci.* 24:9760-9769.

98. Crino, P., K. Khodakhah, K. Becker, S. Ginsberg, S. Hemby, and J. Eberwine. 1998. Presence and phosphorylation of transcription factors in developing dendrites. *Proc. Natl. Acad. Sci. USA* 95:2313-2318.

99. Karpova, A., P. P. Sanna, and T. Behnisch. 2005. Involvement of multiple phosphatidylinositol-3-kinase-dependent pathways in the persistence of late-phase long term potentiation expression. *Neuroscience*, E-pub ahead of print.

100. Pittenger, C., and E. R. Kandel. 2003. In search of general mechanisms for long-lasting plasticity: *Aplysia* and the hippocampus. *Philos. Trans. R. Soc. Lond. B* 358:757-763.

101. Müller, U., and T. J. Carew. 1998. Serotonin induces temporally and mechanistically distinct phases of persistent PKA activity in *Aplysia* sensory neurons. *Neuron* 21:1423-1434.

102. Michael, D., K. C. Martin, R. Seger, N. M. Ning, R. Baston, and E. R. Kandel. 1998. Repeated pulses of serotonin required for long-term facilitation activate mitogen-activated protein kinase in sensory neurons of *Aplysia*. *Proc. Natl. Acad. Sci. USA* 95:1864-1869

103. Martin, K. C., D. Michael, J. C. Rose, M. Barad, A. Casadio, H. Zhu, and E. R. Kandel. 1997. MAP kinase translocates into the nucleus of the presynaptic cell and is required for long-term facilitation in *Aplysia*. *Neuron* 18:899-912.

104. Mauelshagen, J., G. R. Parker, and T. J. Carew. 1996. Dynamics of induction and expression of long-term synaptic facilitation in *Aplysia*. *J. Neurosci.* 16:7099-7108.

105. Martin, K. C. 2002. Synaptic tagging during synapse-specific long-term facilitation of *Aplysia* sensory-motor neurons. *Neurobiol. Learn. Mem.* 78:489-497.

106. Khan, A., A. M. Pepio, and W. M. Sossin. 2001. Serotonin activates S6 kinase in a rapamycin-sensitive manner in *Aplysia* synaptosomes. *J. Neurosci.* 21:382-391.

# TABLE 1

# Standard Model Parameter Values

| Parameters and Values | Biochemical Significance |
|---|---|
| $k_{act1} = 200$ μM min$^{-1}$, $k_{deact1} = 1.0$ min$^{-1}$, $k_{act2} = 2.5$ μM min$^{-1}$, $k_{deact2} = 5.0$ min$^{-1}$, $k_{act3} = 10.0$ min$^{-1}$, $k_{deact3} = 0.05$ min$^{-1}$, $K_{syn} = 0.7$ μM, $K_{nuc} = 0.3$ μM, $\tau_{PKA} = 15$ min, $K_{camp} = 0.5$ μM | Rate constants, Michaelis constants, and Hill coefficients for activation of CaM kinases (Eqs. 1-3). Parameters for PKA activation (Eq. 4). |
| $[Raf]_{tot} = [MAPKK]_{tot} = [MAPK]_{tot} = 0.25$ μM, $k_{f,Raf}$ (basal) $= 0.0075$ min$^{-1}$, $k_{b,Raf} = 0.12$ min$^{-1}$, $k_{f,MAPKK} = 0.6$ min$^{-1}$, $k_{b,MAPKK} = 0.025$ μM min$^{-1}$, $k_{f,MAPK} = 0.52$ min$^{-1}$, $k_{b,MAPK} = 0.025$ μM min$^{-1}$, $K_{MK} = 0.25$ μM, $K_{MKK} = 0.25$ μM, $k_{nuc} = 100.0$ μM$^{-1}$ min$^{-1}$, $k_{cyt} = 2.5$ min$^{-1}$ | Rate constants, Michaelis constants, and conserved total kinase amounts for MAPK cascade activation (Eqs. 5-13) and MAPK nuclear transport (Eqs. 13-14) |
| $k_{phos1} = 0.05$ μM$^{-1}$ min$^{-1}$, $k_{deph1} = 0.02$ min$^{-1}$, $k_{phos2} = 2.0$ μM$^{-1}$ min$^{-1}$, $k_{deph2} = 0.2$ min$^{-1}$, $k_{phos3} = 0.06$ μM$^{-1}$ min$^{-1}$, $k_{deph3} = 0.017$ min$^{-1}$, $k_{phos4} = 0.12$ μM$^{-1}$ min$^{-1}$, $k_{deph4} = 0.03$ min$^{-1}$, $k_{phos5} = 4.0$ μM$^{-1}$ min$^{-1}$, $k_{deph5} = 0.1$ min$^{-1}$ | Rate constants for phosphorylation and dephosphorylation of synaptic tag substrates and transcription factors (Eq. 16) |
| $k_{syn} = 1.0$ μM min$^{-1}$, $k_{synbas} = 0.0004$ μM min$^{-1}$, $k_{deg} = 0.01$ min$^{-1}$, $K_1 = K_2 = 1.0$ | Rate constants and Michaelis constants for GPROD synthesis and degradation (Eq. 17). |
| $k_W = 2.0$ μM$^{-1}$ min$^{-1}$, $\tau_W = 140{,}000$ min, $K_P = 0.03$ μM, $V_P = 0.0003$ μM$^{-1}$ min$^{-1}$, $\tau_P = 1{,}000$ min | Rate constants and time constants for changes in synaptic weight, and parameters for [P] dynamics (Eqs. 18-19). |

## FIGURE LEGENDS

FIGURE 1. Schematic of the model. Synaptic stimulation elevates $Ca^{2+}$ and cAMP and activates the MAPK cascade. $Ca^{2+}$ activates CAMKII and CAM kinase kinase (CAMKK). CAMKK and $Ca^{2+}$ activate CAMKIV. cAMP activates PKA. Activated MAPK, PKA, and CAMKII phosphorylate synaptic substrates (Tag-1 – Tag-3). CAMKIV and MAPK phosphorylate transcription factors (TF-1, TF-2). A variable TAG, denoting the synaptic "tag" needed for L-LTP, equals the product of the fractional phosphorylations of Tag-1 – Tag-3. For L-LTP induction, a gene product must be induced. Induction requires phosphorylation of TF-1 and TF-2. L-LTP induction is modeled as an increase in a synaptic weight W. The rate of increase is proportional to the value of the synaptic tag and to the amount of gene product.

FIGURE 2. Simulations of L-LTP inducing protocols. (A) Tetanic protocol. Each of three tetani briefly elevates $[Ca^{2+}]$, $[cAMP]$, and the rate constant $k_{f,Raf}$ for Raf activation. The red bar represents concurrent elevations in both cytosolic and nuclear $[Ca^{2+}]$. (B) Theta-burst protocol, simulated with a single brief increase in $[Ca^{2+}_{syn}]$ and $[Ca^{2+}_{nuc}]$, $[cAMP]$, and $k_{f,Raf}$. The elevations in $[cAMP]$ and $k_{f,Raf}$ are larger than with the tetanic protocol. For this and the other protocols, the relative heights of the red, green, and blue bars qualitatively reflect the differing amplitudes of the $[Ca^{2+}]$, $[cAMP]$, and $k_{f,Raf}$ elevations, respectively. (C) Pairing protocol. Sixty short bursts of action potentials are each simulated with a relatively small, brief increase in $[Ca^{2+}_{syn}]$ and $[Ca^{2+}_{nuc}]$. $[cAMP]$ and $k_{f,Raf}$ are elevated during the 5-min protocol and for 1 min afterwards. (D) "Chem-LTP". During a 30-min interval, $[Ca^{2+}_{syn}]$ and $[Ca^{2+}_{nuc}]$ are slightly elevated, whereas $[cAMP]$ and $k_{f,Raf}$ are elevated more than in any other protocol.

FIGURE 3. (A) Changes in active CAMKII, active CAMKIV, and active Raf during and after three simulated tetanic stimuli. (A)-(D) use the same stimulus protocol. (B) Changes in synaptic and somatic active MAPK, nuclear active MAPK, and the synaptic tag. (C) Changes in the synaptic tag and the gene product assumed necessary for L-LTP. (D) Changes in the synaptic weight variable W and the concentration of the precursor protein P. For plotting, time courses were vertically scaled (but not horizontally scaled) as described in Numerical Methods. In Figs. 3-7, the variables representing enzyme concentrations and the variables $[P]$ and $[GPROD]$ have units of $\mu M$. The other variables, such as W and TAG, are non-dimensional.

FIGURE 4. Time courses of the synaptic weight W following the tetanic protocol of Fig. 3. Four cases are simulated: 1) no kinase inhibition (control), 2) the concentration of active CAMKII in Eq. 16 is scaled down by 90% during the tetanic stimulation and for 50 min after (-CAMKII), 3) the concentration of active CAMKIV is scaled down by 90% during and at all times after stimulation (-CAMKIV), 4) MAPKK activation (the rate constant $k_{f,MAPKK}$) is inhibited by 90% during stimulation and for 10 min after (-MAPKK). For the cases of CAMKII and CAMKIV inhibition, the W time courses are virtually identical.

FIGURE 5. (A) Changes of the synaptic weight W following four stimulus protocols: 1) the tetanic stimuli used in Fig. 3, 2) a theta-burst stimulus protocol (TBS), 3) application of chemical for 30 min (Chem), and 4) a paired stimulus protocol (53). (B) Changes of $[MAPK_{act}]$, [GPROD], the synaptic tag, and W during and after the paired stimulus protocol.

FIGURE 6. (A) Changes of $[MAPK_{act}]$, the synaptic tag, and active CAMKII during and after a simulated 30-min chemical application. (B) Changes of [GPROD] and W. Also shown is the attenuated W time course (-MAPKK) observed when $k_{f, MAPKK}$ is reduced by 90% during and for 5 min after the chemical application.

FIGURE 7. (A) Schematic of the simulation of synaptic tagging. Three tetani, identical to Fig. 2A except with an inter-stimulus interval of 10 min, are applied to synapse A. One hour after the first tetanus to synapse A, synapse B is likewise given 3 tetani. Only the tetani to synapse A activate gene expression (GPROD synthesis). (B) Time courses of the tag at synapse A, the tag at synapse B, [GPROD], and W for synapses A and B. The W (tetanic) time course represents L-LTP of synapse A, the W (tagged) time course represents L-LTP of synapse B. Synthesis of GPROD is blocked 35 min after the first set of tetani by setting the rate constants $k_{syn}$ and $k_{synbas}$ (Eq. 17) to zero.

Smolen et al.,
Figure 1

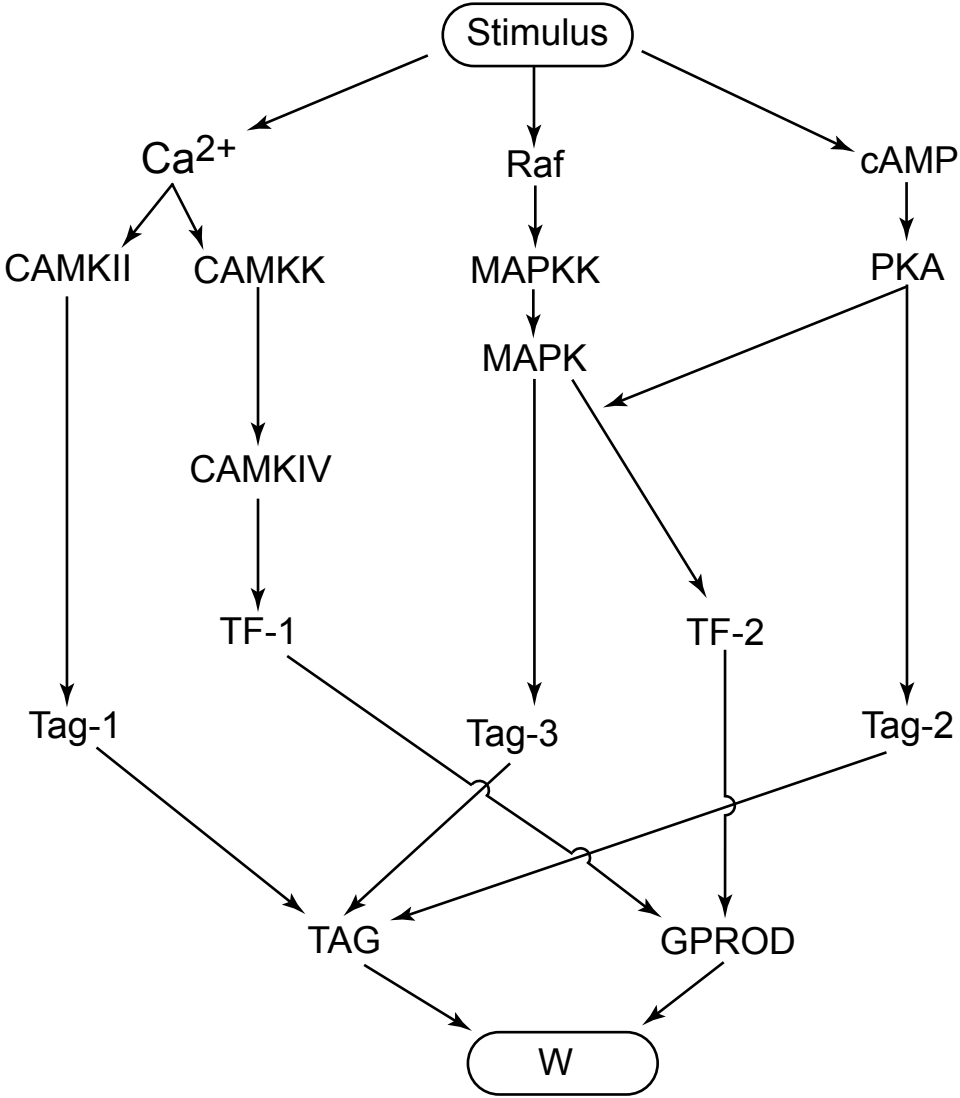

Smolen et al.,
Figure 2

**A. Tetanic L-LTP induction protocol**

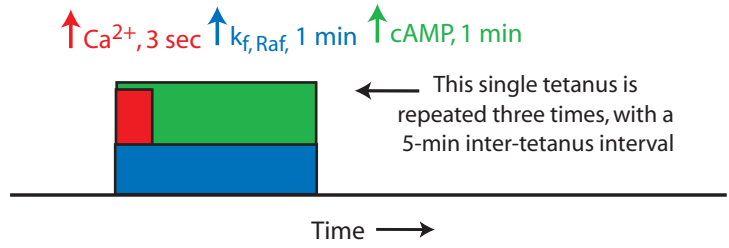

↑ Ca²⁺, 3 sec    ↑ $k_{f, Raf}$, 1 min    ↑ cAMP, 1 min

← This single tetanus is repeated three times, with a 5-min inter-tetanus interval

Time ⟶

**B. Theta-burst induction protocol**

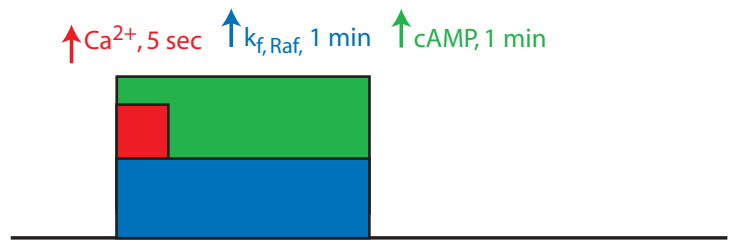

↑ Ca²⁺, 5 sec    ↑ $k_{f, Raf}$, 1 min    ↑ cAMP, 1 min

**C. Heterosynaptic induction protocol**

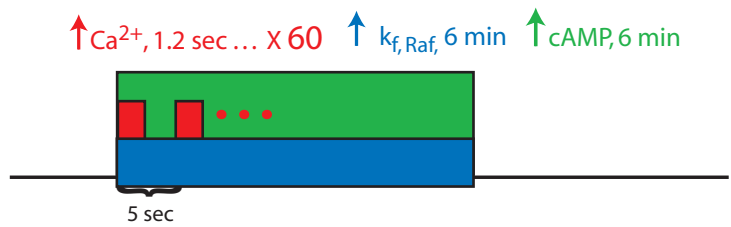

↑ Ca²⁺, 1.2 sec ... X 60    ↑ $k_{f, Raf}$, 6 min    ↑ cAMP, 6 min

5 sec

**D. Chemical induction protocol**

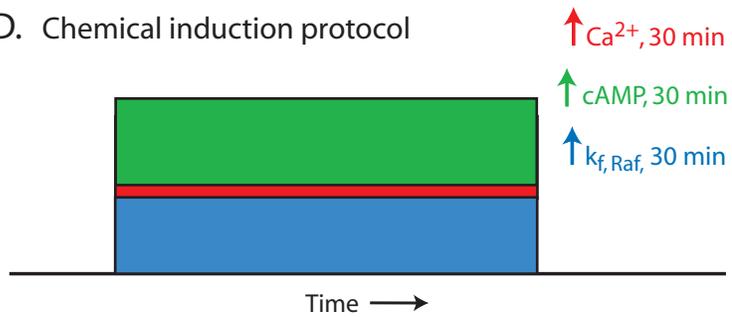

↑ Ca²⁺, 30 min

↑ cAMP, 30 min

↑ $k_{f, Raf}$, 30 min

Time ⟶

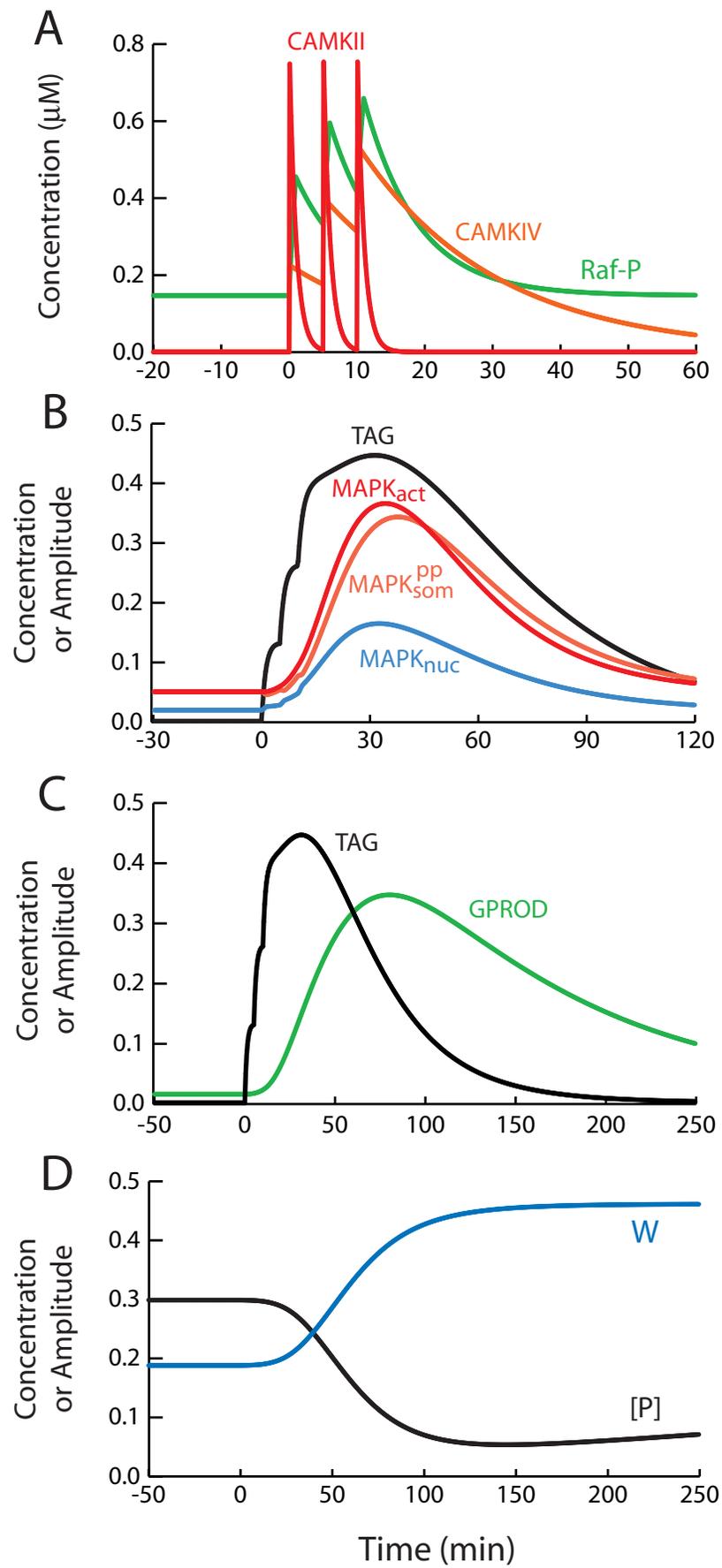

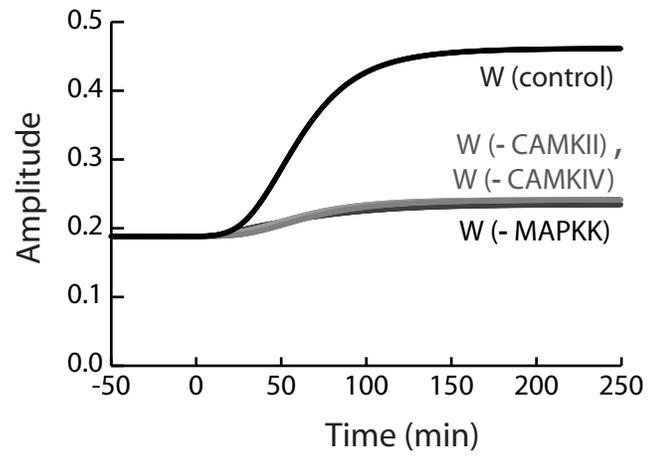

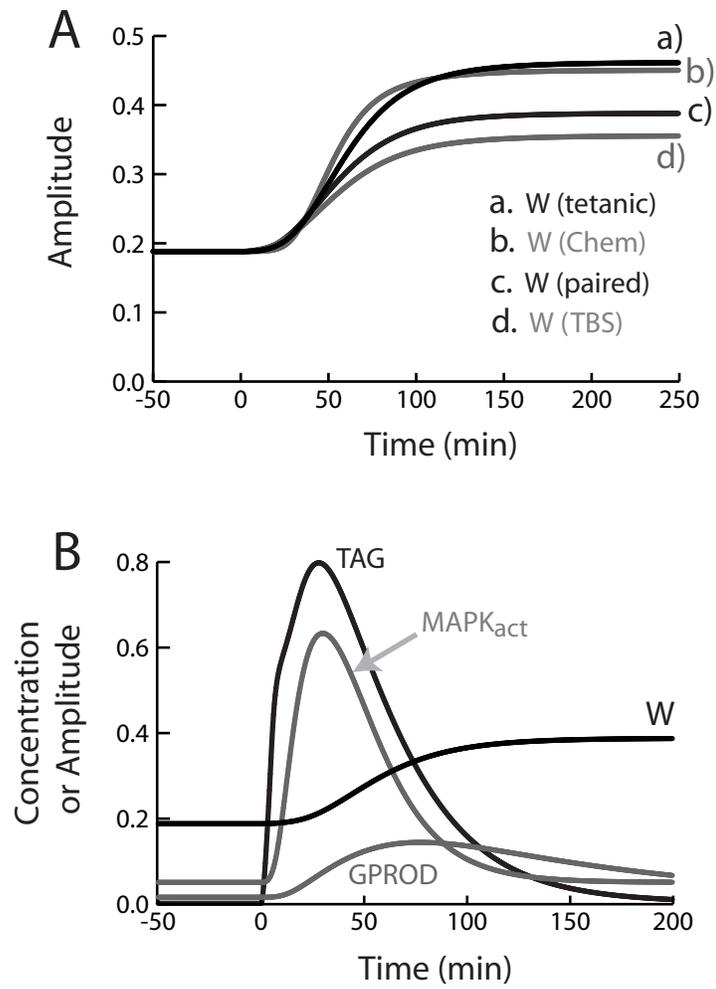

A

Amplitude

a. W (tetanic)
b. W (Chem)
c. W (paired)
d. W (TBS)

Time (min)

B

Concentration
or Amplitude

TAG

MAPK_act

W

GPROD

Time (min)

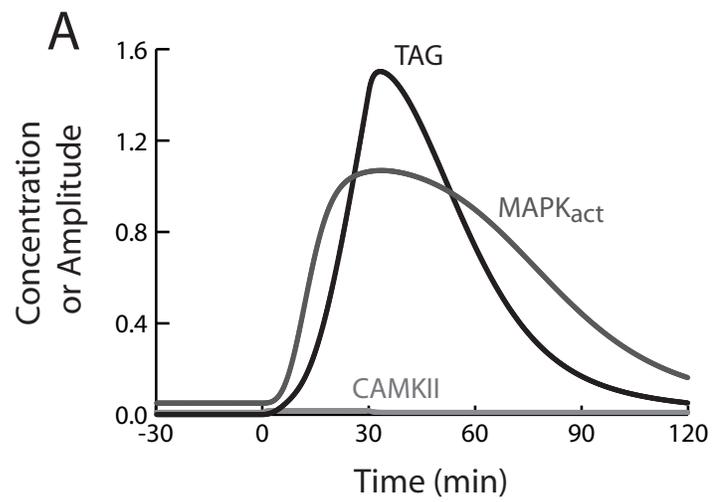

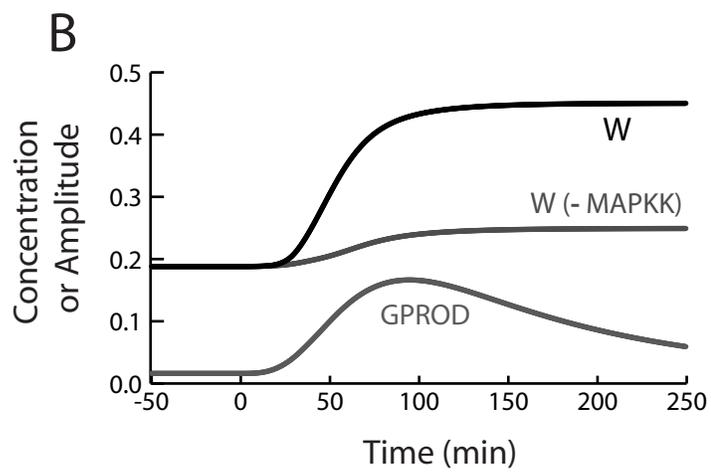

A    Synaptic tagging induction protocol

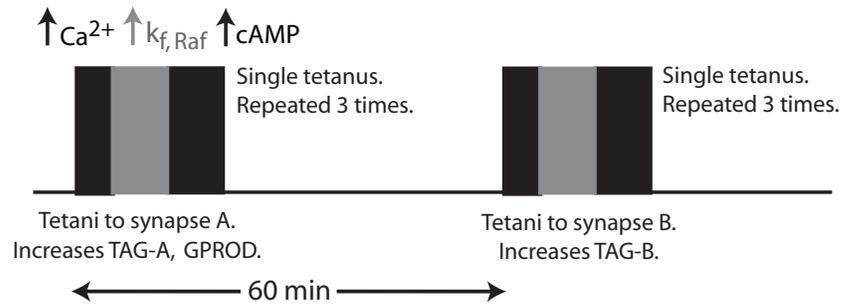

B

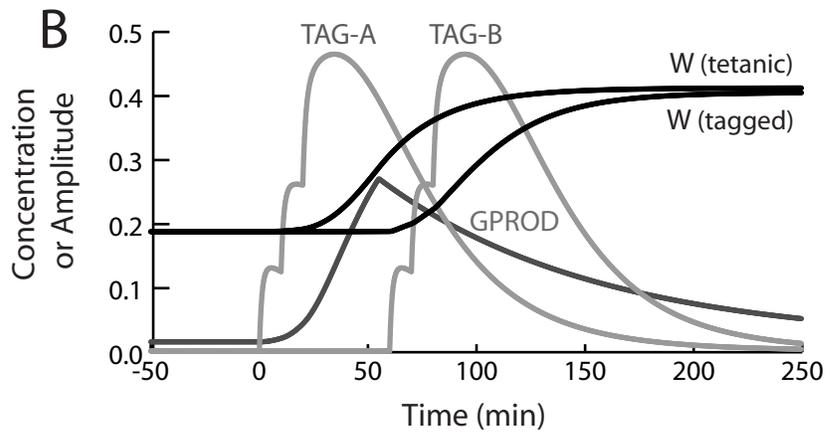